\documentclass[twocolumn,prl,sort&compress,floatfix]{revtex4-1}
\usepackage{amssymb}
\usepackage{amsmath}
\usepackage{mathrsfs}
\usepackage{bm}
\usepackage[dvipsnames]{xcolor}
\usepackage[sort&compress]{natbib}

\usepackage{graphicx}

\usepackage{color}
\definecolor{red}{rgb}{1,0,0}
\definecolor{blue}{rgb}{0,0,1}

\usepackage{xcolor}
\definecolor{royalblue}{rgb}{0.2549,0.3,0.95}

\definecolor{englishred}{rgb}{0.83,0.24,0.1}

\definecolor{carrot}{rgb}{0.93,0.1,0.2}

\begin{document}

\title{Synergy of Topoisomerase and Structural-Maintenance-of-Chromosomes Proteins Creates a Universal Pathway to Simplify Genome Topology}

\author{Enzo Orlandini$^a$, Davide Marenduzzo$^b$, Davide Michieletto$^b$}

\affiliation{$^a$Dipartimento di Fisica e Astronomia ``Galileo Galilei'', sezione
INFN, Universit\`a degli Studi di Padova, via Marzolo 8, I-35131 Padova, Italy. \\
$^b$SUPA, School of Physics and Astronomy, University of 
	Edinburgh, Peter Guthrie Tait Road, Edinburgh, EH9 3FD, UK}

\begin{abstract}
\textbf{Topological entanglements severely interfere with important biological processes. For this reason, genomes must be kept unknotted and unlinked during most of a cell cycle. Type II Topoisomerase (TopoII) enzymes play an important role in this process but the precise mechanisms yielding systematic disentanglement of DNA {\it in vivo} are not clear. Here we report computational evidence that Structural Maintenance of Chromosomes (SMC) proteins -- such as cohesins and condensins -- can cooperate with TopoII to establish a synergistic mechanism to resolve topological entanglements. SMC-driven loop extrusion (or diffusion) induces the spatial localisation of essential crossings in turn catalysing the simplification of knots and links by TopoII enzymes even in crowded and confined conditions. The mechanism we uncover is universal in that it does not qualitatively depend on the specific substrate, whether DNA or chromatin, or on SMC processivity; we thus argue that this synergy may be at work across organisms and throughout the cell cycle.}
\end{abstract}

\maketitle

Genomes are long polymers stored in extremely crowded and confined environments; the ensuing inevitable entanglements are thought to cause DNA damage, interfere with gene transcription, DNA replication and interrupt anaphase, eventually leading to cell death~\cite{Bates2005,Grosberg1993a,Duplantier1995}. 
{\it In vitro} and under dilute conditions, TopoII proteins efficiently resolve topological entanglements and stabilise a population of knotted DNA below the expected value in thermodynamic equilibrium~\cite{Rybenkov1997}. These findings can be partially explained by a model where TopoII enzymes recognise specific DNA-DNA juxtapositions~\cite{Vologodskii2001,Yan1999,Liu2010}. Yet, how this model can lead to efficient unknotting and unlinking in crowded environments and crumpled DNA or chromatin substrates is unclear~\cite{Arsuaga2002,MartinezGarcia2014,Grosberg1993a}. 
Even more intriguing is the {\it in vitro} experimental finding that, in presence of polycations~\cite{Krasnow1982} or with superstochiometric abundance of TopoII~\cite{Wasserman1991}, the action of these proteins may {\it increase} the topological complexity of DNA substrates~\cite{Hsieh1980,Krasnow1982,Hsieh1983}.

While it has been suggested that DNA supercoiling may provide a solution for this problem by promoting hooked DNA juxtapositions ~\cite{Vologodskii1996,Witz2011,Racko2015}, this argument is valid only for naked, highly supercoiled DNA, such as bacterial plasmids. The understanding of how efficient topological simplification is achieved in eukaryotes where the genome is packaged into chromatin remains, on the other hand, an outstanding and unresolved problem~\cite{Valdes2018,Bates2005}. 

Here we propose a novel mechanism for efficient topological simplification in DNA and chromatin {\it in vivo} that is based on the synergistic action of SMC-driven loop extrusion~\cite{Nasmyth2001,Alipour2012,Fudenberg2016,Uhlmann2016} (or diffusion~\cite{Brackley2017prl}) and TopoII. We show that the sliding of slip-link-like proteins along DNA and chromatin is sufficient to localise any knotted and linked regions or their essential crossings, in turn catalysing their topological simplification. Our simulations reveal that this mechanism is independent of either substrate condensation or crowding, and is therefore likely to lead to unknotting and unlinking even under extreme conditions such as those in the cell nucleus.
Finally, we discuss our model in the context of recent experiments reporting that SMC proteins are essential to achieve correct sister chromatid decatenation in metaphase~\cite{Piskadlo2017a}, that DNA damage is frequently found in front of cohesin motion~\cite{Vian2018} and that there is a remarkable low frequency of knots in intracellular chromatin~\cite{Valdes2018}.

\begin{figure*}[t!]
	\centering
	\includegraphics[width=1\textwidth]{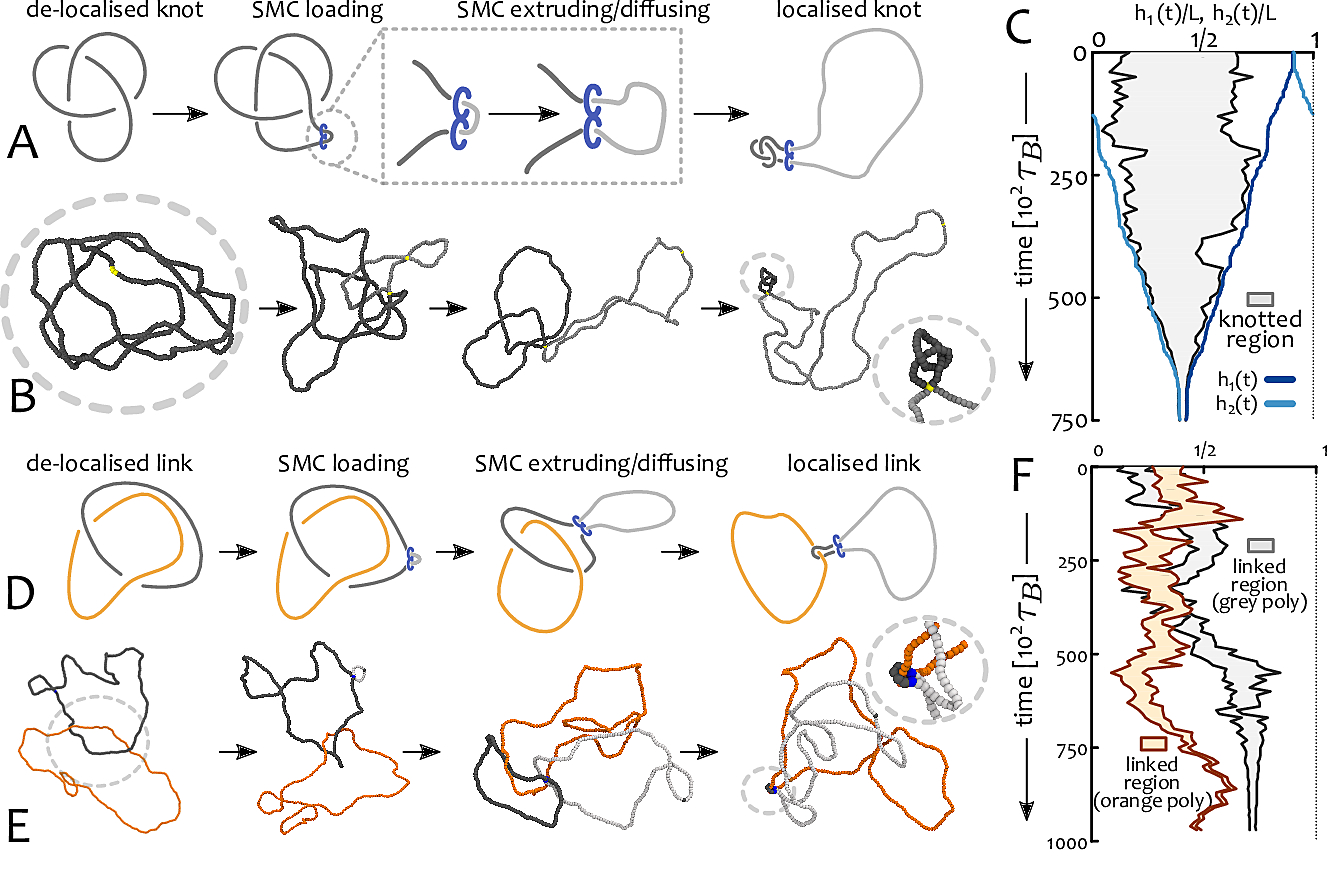}
	\vspace*{-0.7 cm}
	\caption{\textbf{Sliding of SMC proteins Localises Topological Entanglements.} (A) Schematics of knot localisation starting from a fully delocalised trefoil via loop extrusion/diffusion. (B) Corresponding Brownian Dynamics simulations. (C) Kymograph showing the shortest knotted arc along the chain as a function of time. The blue curves show the position of the SMC heads ($h_1(t), h_2(t)$) and demonstrate that the knot localises over time. (D) Schematics of link localisation starting from a delocalised Hopf link. (E) Corresponding Brownian Dynamics simulations. (F) Kymograph showing the shortest linked segments for the two polymers. As the SMC protein is loaded on the grey polymer, the linked region in the sister strand is free to slide and this gives rise to a localised but fluctuating orange-shaded area in the figure. See Supplementary Movies 1 and 2. }
	\vspace*{-0.5 cm}
	\label{fig:panel1}
\end{figure*}

\vspace*{-0.2 cm}
\section*{Results and Discussion}

\vspace*{-0.2 cm}
\subsection*{Model and System Set Up}
We perform Brownian Dynamics (BD) simulations of a generic polymer substrate modelled as a semi-flexbile bead-spring circular chain of $500$ beads of size $\sigma$, taken to be 2.5 nm for DNA~\cite{Rybenkov1993} and 10 nm for chromatin~\cite{DekkerJob2002}. %This model captures generic physical properties of a genome~\cite{Rosa2008} and 
We consider circular chains as representative of DNA plasmids or stably looped genomic regions such as the so-called ``Topologically Associated Domains'' (TADs) bound by CTCF proteins~\cite{Rao2014} and knotted and linked topologies as capturing topological entanglements that typically occur in genetic materials~\cite{Sogo1999,Arsuaga2002,Trigueros2007,Marenduzzo2009,Valdes2018} (see Fig.~\ref{fig:panel1}). %This choice also simplifies the computational task of identifying topological changes of the substrate but, as we discuss below, this is not a crucial requirement for the proposed mechanism to work. 
Unlike previous works~\cite{Rosa2008,Goloborodko2016}, here we explicitly forbid spontaneous strand-crossing events by imposing that any pair of consecutive beads are connected by finitely extensible (FENE) springs~\cite{Kremer1990} while non-consecutive ones are subject to a purely repulsive (WCA) potential. A Kratky-Porod term is used to set up the persistence length at $l_p=20\sigma$. Note, however, that the results are not qualitatively affected by this choice (SI Appendix).

\vspace{-0.25 cm}
\subsection*{A Slip-Link Model for SMC} 
SMC proteins, including condensin and cohesin, are thought to regulate genome architecture across organisms by topologically embracing DNA or chromatin in a slip-link-like fashion~\cite{Hirano2016,Nasmyth2001,Gruber2003,Uhlmann2016,Wilhelm2015}. Recent experiments \emph{in vitro} suggest that condensin can move directionally at a speed $v \simeq 0.6-1.5 \, kb/s$~\cite{Ganji2018} and that cohesin performs diffusive sliding with diffusion constant $D \simeq 0.1 - 1 \, \mu m^2/s$~\cite{Kanke2016,Stigler2016}. Previous work has crudely modelled SMC proteins as harmonic springs between non-consecutive chromosome segments which were dynamically updated (irrespectively of local constraints) in order to extrude loops~\cite{Fudenberg2016,Goloborodko2016,Sanborn2015}. In contrast, here we account for both the steric hindrance and the slip-link nature of the SMC complex by modelling the SMC bond with a finitely extensible (FENE) spring so that it is energetically very unfavourable for a third segment to cross through the gap in between the bonded beads. The two chromosome segments bound by the SMC protein at time $t$, or SMC ``heads'', are denoted as $h_1(t)$ and $h_2(t)$ and updated at rate $\kappa$ (SI Appendix). We here focus on processive complexes and thus update the location of the heads as $h_1(t+dt)=h_1(t)+1$ and $h_2(t+dt)=h_2(t)-1$ only if the Euclidean distance between the next pair of beads is shorter than $1.3\sigma$. 
This rule ensures that no third bead can pass through the segments bonded by the SMC protein during the update step and it effectively slows down the processivity of the complex depending on the instantaneous substrate conformation. 
We highlight that the speed of the extrusion process does not qualitatively affect the synergistic mechanism found here, only its overall completion time.

% from $v = 2\sigma \kappa = 2$ $10^{-3} \sigma/\tau_B$ to about $v \simeq 5 \, 10^{-5} \sigma/\tau_B$ or $v_{30nm}=0.125$ nm/s $=1.5$ kb/min assuming a chromatin compaction of 100 bp/nm~\cite{Brackley2017prl}.

%\dmi{Finally, it is known that SMC proteins display a finite unloading time (about $20$ minutes for cohesin~\cite{Busslinger2017}), yet here we will assume this time infinite. This is justified by the regime of polymer lengths considered here: this maps to large TADs \emph{in vivo} and given the speed/diffusion found \emph{in vitro}, these can be comfortably fully covered before unloading from the substrate (SI Appendix for dynamically loaded SMCs). }

\vspace{-0.25 cm}
\subsection*{SMC Sliding Localises Topological Entanglements}
Thermally equilibrated knotted or linked polymers in good solvent display weakly localised topological entanglements~\cite{Orlandini2009a,Grosberg2007,Caraglio_et_al_Sci_Rep_2017}, i.e. the shortest arc that can be defined knotted or linked, $l_K$, grows sublinearly with the overall contour length $L$, as $l_K \sim L^{0.75}$ (see Fig.~\ref{fig:panel1}A)~\cite{Marcone2005,Tubiana2011knots,Caraglio_et_al_Sci_Rep_2017}. 
Further topological de-localisation is achieved by isotropic confinement~\cite{Tubiana2011prl,Marenduzzo2013} and crowding~\cite{DAdamo2015}, both conditions that are typically found \emph{in vivo}. Since de-localisation of essential crossings is likely to hinder TopoII-mediated topological simplification, it is natural to ask if there exists a physiological mechanism that counteracts topological de-localisation {\it in vivo}.

To address this question we performed BD simulations of directed loop extrusion on thermalised polymers which display de-localised entanglements (Fig.~\ref{fig:panel1}A). The ensuing extrusion, or growth, of the subtended loop can be monitored by tracking the location of the SMC heads $h_1(t)$ and $h_2(t)$ (see blue curves in Fig.~\ref{fig:panel1}C). At the same time, we used well-established existing algorithms~\cite{Tubiana2011knots,Tubiana2011prl} (publicly accessible through the server \url{http://kymoknot.sissa.it}~\cite{Tubiana2018}) to compute the shortest portion of the chain hosting the knot. 
%[Note that this is a ``physical knot'' as a closure procedure is required before its identification through topological invariants.] 
We observed that the shortest knotted arc $l_K$, initially spanning a large portion of the polymer, progressively shrinks into a region whose boundaries match the location of the SMC heads. Notably, in the large time limit, all the essential crossings forming the knot (in Fig,~\ref{fig:panel1} a trefoil, $3_1$) were observed to be localised within a segment $l \ll L$ (see Fig.~\ref{fig:panel1}B). A similar localisation effect could be achieved on a pair of linked polymers (see Fig.~\ref{fig:panel1}D-F). 

%% DIFFUSION DISCUSS LATER! 
%It is important to stress that the discovered SMC-mediated localisation effect is not strictly due to the processivity of the protein and may also be achieved even if SMC slides diffusively along DNA~\cite{Brackley2017prl}. To localise an entanglement via diffusing slip-link-like elements one would require a number of SMC proteins that increases with the entropic pressure of the knot (or link) which needs to be overcome~\cite{Zandi2003}. Additionally, we highlight that a mechanism of ratcheted diffusion with specific loading sites for the SMC protein (such as NIPBL for the SMC cohesin complex~\cite{Zuin2014}) would further enhance the efficiency of topological localisation with respect to diffusive SMC loaded randomly along the fibre~\cite{Brackley2017prl}.
%%

Importantly, SMC-driven topological localisation does not require a topologically closed (circular) substrate to function. Physiologically occurring loops, e.g. between enhancer and promoters~\cite{Alberts2014}, CTCFs at TAD boundaries~\cite{Rao2014} or protein bridges~\cite{Brackley2016nar}, define transient and stably-looped genomic regions which would effectively act as circular substrates and entrap topological entanglements such as knots and links.

%\begin{figure}[t!]
%	\centering
%	\includegraphics[width=0.45\textwidth]{panel2.jpg}
%	\caption{Knot localisation by SMC in transiently looped regions.}
%	\label{fig:panel2-looped-bridges}
%\end{figure}

\vspace{-0.2 cm}
\subsection*{A model for SMC-recruited TopoII}
\begin{figure}[t!]
	\centering
	\includegraphics[width=0.45\textwidth]{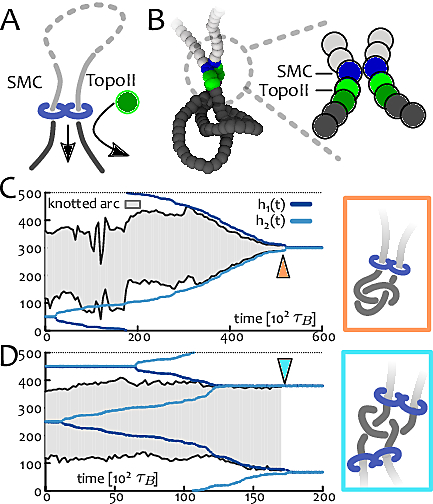}
	\vspace*{-0.3 cm}
	\caption{\textbf{SMC-recruited TopoII.} (A) Motivated by experimental findings~\cite{Uuskula-Reimand2016,Vian2018}, we assume that TopoII is co-localised with SMC and it is found on the outside of the SMC-mediated loop (dark segments). (B) Implementation of (A) in a bead-spring polymer model: the SMC slip-link is enforced by a FENE bond between blue beads which are updated in time. TopoII beads (green) are set to display a soft repulsive potential with all other beads thus allowing thermally-activated strand crossing. Dark and light green beads have different energy barriers against overlapping ($5 k_BT$ and $20 k_BT$, respectively). (C-D) Kymographs showing synergistic knot simplification:  in (C), SMC-driven loop-extrusion localises the shortest knotted arc while in (D), two SMC localise only the knot's essential crossings (see insets). We find that (D) is predominant for diffusive SMC (SI Appendix). See also Suppl. Movies 3 and 7.
	%To determine the location and bounds of the knot we employed the algorithm described in Refs.~\cite{Tubiana2011knots,Tubiana2011prl,Tubiana2018}. A running average over 5 timesteps was employed to ease visualisation.  \dmi{(D) The action of multiple SMC (2 in the figure) can lead to conformations in which the  whereas its contour length is not (cyan in the inset). These situations can still be simplified and reduced to simpler knots (here the $3_1$ is sent to the unknot). We find that this pathway is predominant in the case of diffusive SMC (SI Appendix). }  
    }
	\label{fig:smc-topo-model}
	\vspace*{-0.4 cm}
\end{figure}

Having shown that SMC complexes can induce the {\it localisation} of topological entanglements, we next asked whether downstream action of TopoII on localised entanglements could provide a fast and efficient mechanism for topological {\it simplification}. To this end, here we propose a model in which TopoII is directly recruited by SMC (Fig.~\ref{fig:smc-topo-model}A) and is motivated by recent experiments reporting direct interaction between TopoII and SMC cohesin \emph{in vivo}~\cite{Uuskula-Reimand2016,Vian2018}. Our model is qualitatively different from random passage models for TopoII~\cite{Flammini2004,Goloborodko2016,Michieletto2014a} and it is practically implemented by allowing only the two nearest beads in front of the ones forming the SMC heads, i.e. $h_{1/2}(t) \pm 1$, $h_{1/2}(t) \pm 2$, to undergo strand-crossing (SI Appendix and Fig.~\ref{fig:smc-topo-model}B).
\vspace{-0.3 cm}

\begin{figure*}[t!]
	\centering
	\vspace*{-0.3 cm}
	\includegraphics[width=1\textwidth]{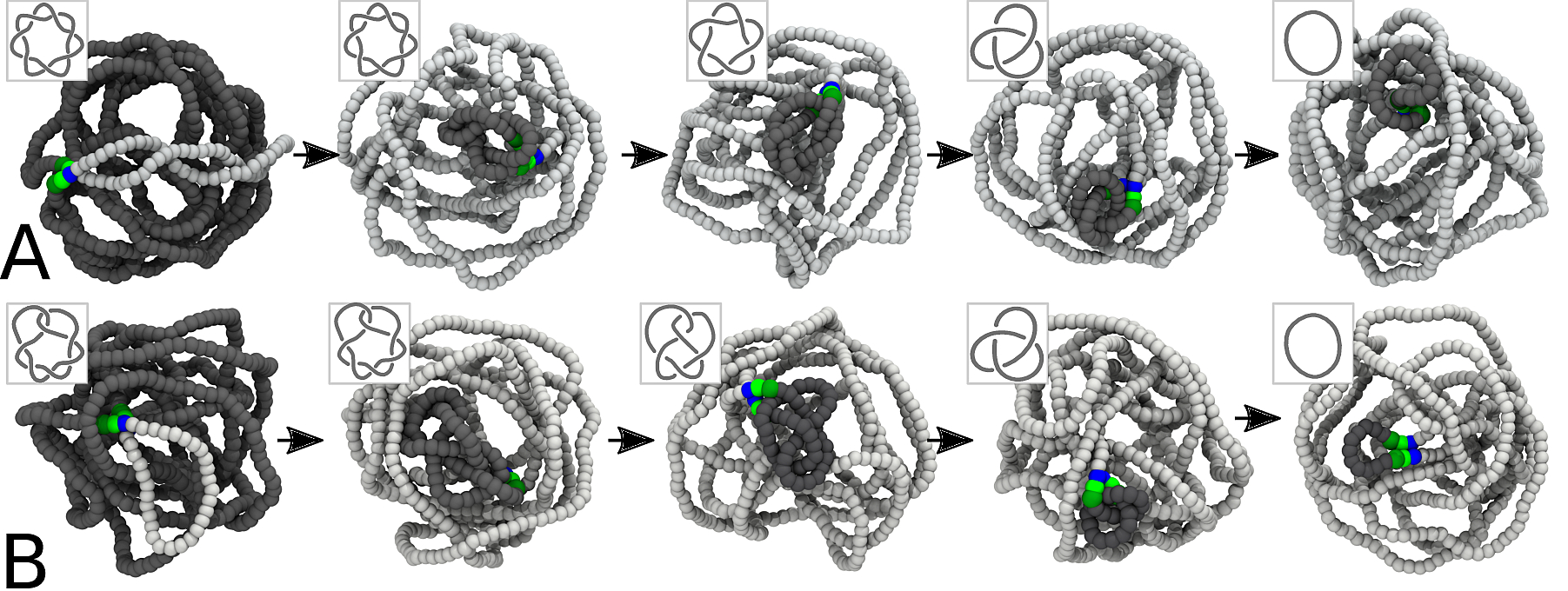}
	\vspace*{-0.6 cm}
	\caption{\textbf{Efficient Unknotting under Confinement.} The synergistic action of SMC and TopoII proteins can systematically simplify knotted substrates even under confinement. Here we show the case of a torus ($7_1$) and a twist ($7_2$) knots confined within a sphere with radius $R_c/\langle R_g \rangle \simeq 1/3$. In the snapshots, light-grey beads are the ones that have been extruded by, hence behind, the SMC. Dark-grey beads are the ones outside the extruded loop. Blue beads mark the location of the SMC heads. Green and dark-green beads mark the location of TopoII, as described in the text.  (A) Unknotting of a $7_1$ knot through the ``cascade'' of torus knots $5_1$ and $3_1$. (B) Unknotting of a $7_2$ knot through $5_2$ and $3_1$ knots. Direct simplification $7_2 \rightarrow 0_1$ is also observed in more than half of the simulations (see Table T\ref{table:tab} and SI Appendix).  See Suppl. Movies 4 and 5.
		%	 (where we show $7_1$ and $7_2$ wihtin a sphere of radius $R/\langle R_g \rangle \simeq 3$) 
	}
	\vspace*{-0.4 cm}
	\label{fig:smc-topo-confined}
\end{figure*}

\subsection*{Localising Topological Entanglements Catalyses TopoII-mediated Simplification}
We first tested whether the local recruitment of TopoII by SMC can efficiently simplify the substrate topology. To this end, we performed BD simulations initialised from equilibrated configurations containing a de-localised trefoil knot ($3_1$) and loaded one SMC protein recruiting a TopoII enzyme, as discussed in the previous section (see Fig.~\ref{fig:smc-topo-model}). We monitored the time evolution of the substrate topology by computing its instantaneous Alexander polynomial~\cite{Tubiana2018} while tracking both the position of the SMC heads and the boundaries of the knotted region~\cite{Tubiana2011knots}. Remarkably, in all the independent replicas of the system, the synergy of SMC and TopoII was able to simplify the topology of the substrate down to the unknot (Fig.~\ref{fig:smc-topo-model}). Importantly, the topological simplification occurred only after the knot localisation by the single SMC protein (see Fig.~\ref{fig:smc-topo-model}C). To explain this finding one may argue that a localised knot enhances intra-knot contacts over ones occurring between any other two segments of the polymer; in turn, this conformational bias favours the crossing of intra-knot segments and catalyses the decrease in topological complexity. Equivalently, one may recall that the probability of finding an unknot in equilibrium is exponentially small with the substrate length $L$, i.e. $P_{0} \sim e^{-L/L_0}$~\cite{Orlandini2007}; inducing knot localisation effectively yields $L<L_0$ thus greatly enlarging the statistical weight of unknotted conformations.  
%In other words, knot (or link) localisation increases the local entropic pressure within a knotted region~\cite{Zandi2003} which then promotes the knot (or link) simplification.
  
By loading more than one SMC proteins onto the substrate we discovered that there exist another pathway for topological simplification. This involves the localisation of the essential crossings but does not lead to a minimal knotted arc $l_K \ll L $; this pathway is selected when a pair of SMCs extrude loops simultaneously from within and outside the knotted region (see Fig.~\ref{fig:smc-topo-model}D) and it yields polymer conformations that are reminiscent of those computationally observed in DNA knot translocation~\cite{Suma2017}. Interestingly, this unknotting pathway is favoured and often observed in simulations of diffusing slip-links (SI Appendix and Suppl. Movie 7). 
  
For simplicity, we assumed an infinitely long residency time of SMC proteins. While a population of condensin is stably bound in mitosis~\cite{Gerlich2006}, cohesin is known to turn-over in about $\tau=20$ minutes through inter-phase~\cite{Busslinger2017}. At a speed $v \simeq 1 \, \rm{kb/s}$~\cite{Ganji2018}, SMC proteins can extrude loops of length  $l = v \tau > \rm{Mb}$ during their lifetime. By diffusing at $D \simeq 0.1 - 1 \, \mu m^2/s$~\cite{Stigler2016,Kanke2016} SMC proteins can cover distances of about $\sqrt{D \tau} \simeq 200-700 \, \rm{kb}$ over a loosely packed chromatin storing 200 bp in 10 nm (SI Appendix). In either cases the processivity ($p=v\tau$~\cite{Fudenberg2016} or $p=\sqrt{D \tau}$~\cite{Brackley2017prl}) of the SMC is comparable (or larger) than both, the length of typical TADs -- which have median $185$ kb in humans~\cite{Rao2017} -- and that of our polymer substrate ($200-500$ kb). In the SI Appendix, we show that when the SMC processivity is shorter than the length of the substrate our synergistic model can still achieve topological simplification, albeit in a stochastic sense.

We finally highlight that the observed topological simplification is different from all existing alternative mechanisms accounting for the action of TopoII alone~\cite{Flammini2004,Vologodskii2009}. Our mechanism also works in the absence of high levels of supercoiling, known to provide another non-equilibrium pathway for post-replicative decatenation~\cite{Racko2015}, but not documented in eukaryotic chromatin. 

\vspace*{-0.2 cm}
\subsection*{Synergistic Topological Simplification is Efficient in Crowded and Confined Conditions}

One of the major problems in elucidating TopoII-mediated topological simplification \emph{in vivo} is that it must ``recognise'' the \emph{global} topology of the substrate while performing \emph{local} strand-crossings. Hooked DNA juxtapositions between pre-bent segments may provide a simple read-out mechanism to simplify localised knots in dilute conditions~\cite{Vologodskii2001,Flammini2004,Burnier2007}. However, this is not a viable pathway in crowded or confined conditions such as those {\it in vivo} because (i) in dense solutions many DNA-DNA juxtapositions occur by random collision regardless of the local bending and (ii) knots and other forms of topological entanglement tend to de-localise under isotropic confinement~\cite{Tubiana2011prl}. It is thus natural to ask whether the synergistic mechanism proposed here may provide a robust pathway to simplify genome topologies under confinement, as required within the nucleus of cells.
To this end we performed simulations on polymers displaying a range of knot types and confined within a sphere of radius $R_c$ about 3 times smaller than the mean gyration radius of the same polymer in equilibrium in good solvent, $\langle R_g \rangle$. Remarkably, we discover that the synergistic action of SMC and TopoII can efficiently simplify the substrate topology even in this extreme confinement regime. In particular, as the SMC protein slides along the crumpled substrate, we observe configurations in which a third segment is found in front of the extruding fork (see Fig.~\ref{fig:smc-topo-confined}), highly reminiscent of hooked juxtapositions~\cite{Vologodskii2009,Burnier2007}. Within our model, these events are spontaneous, in that they are due to the linear reeling in of the substrate through the SMC slip-link. These findings also suggest that the recruitment of TopoII in front of the extruding motion of the SMC~\cite{Uuskula-Reimand2016}, may be an evolutionary optimal strategy to resolve topological entanglements.

A mechanism that can achieve efficient topological simplification under confinement has never been proposed before and our simulations even suggest that our model may be the more efficient the stronger the confinement (SI Appendix). This can be explained as the entropic penalty for forming a loop of size $l$ by the SMC complex scales as $c k_BT \log{l}$ with $c$ the contact exponent~\cite{Gennes1979,Brackley2017prl,Duplantier1989}. Thus, on crumpled substrates, i.e. $c \simeq 1$, the entropic penalty is smaller than on swollen ones $c \simeq 2.1$. This implies that the extrusion/diffusion of the SMC is less hindered under confinement and the localisation of the knot is thus achieved more quickly (SI Appendix, Fig.~S4).

\begin{table}[t!]
	\setlength\tabcolsep{4 pt} % default value: 6pt
	\centering
	\begin{tabular}{c | c c | c c | c } \toprule
		& \multicolumn{2}{c}{Synergistic } & \multicolumn{2}{c}{RP} &  HJ \\ 
		% \hline
		& Free & Confined  & Free  & Confined  & Free  \\ 
		& this work & this work  & Ref.~\cite{Vazquez2007} & this work  & Ref.~\cite{Burnier2007} \\
		\hline
		%%NOW OK FREE AND CONFINED
		$7_1 \rightarrow K$ & $0.02$ & $0.06$ & $0.66$ & $0.98$ & -- \\	
		$7_1 \rightarrow 5_1$ & $0.98$ & $0.92$ & $0.34$ & $0.02$ & -- \\	
		$7_1 \rightarrow 3_1$ & $0$ & $0.02$ & $0$ & $0$ & -- \\	
		\hline
		%% THIS IS OK FREE AND CONFINED
		$5_2 \rightarrow K$ & $0$ & $0.1$ & $0.49$ & $0.8$ & $0.26$ \\ 
		$5_2 \rightarrow 3_1$ & $0.5$ & $0.25$ & $0.2$ & $0.13$ & $0.23$ \\
		$5_2 \rightarrow 0_1$ & $0.5$ & $0.65$ & $0.31$& $0.07$ & $0.51$  \\
		\hline
		$5_1 \rightarrow K$ & $0$ & $0.06$ & $0.69$ & $0.8$ & -- \\ 
		$5_1 \rightarrow 3_1$ & $1$ & $0.94$ & $0.31$ & $0.13$ & -- \\ 
		\hline
		%% NOW OK FREE AND CONFINED
		$4_1 \rightarrow K$ & $0$ & $0.04$ & $0.16$ & $0.84$ & -- \\ 	 
		$4_1 \rightarrow 0_1$ & $1$ & $0.96$ & $0.84$ & $0.16$ & -- \\ 	 
		\hline
		$3_1 \rightarrow K$ & $0$ & $0.15$ & $0.22$ & $0.87$ & $0.2$\\ 	 
		$3_1 \rightarrow 0_1$ & $1$ & $0.85$ & $0.78$ & $0.13$ & $0.8$\\ 	 
		\hline
	\end{tabular}
	\caption{Knot transition probabilities in different models. Topology simplification through the synergistic model proposed in this work is compared with RP (Ref.~\cite{Vazquez2007}) and HJ (Ref.~\cite{Burnier2007}) models. The confined case is compared with RP simulations performed in this work. $\mathcal{K}_1 \rightarrow K$ denotes transition to any knot $K$ with equal or larger minimal crossing number. Full table is given in the SI Appendix.
	%(HJ, hooked juxtaposition; RP, Random Passage). 
	}
	\vspace*{-0.3 cm}
	\label{table:tab}
\end{table}

\vspace*{-0.1 cm}
\subsection*{Comparison of the Synergistic Versus Random Passage and Hooked Juxtaposition Models}
To compare the efficiency of the mechanism proposed here against previous models for TopoII, we estimated the transition probabilities within the space of knots, $P(\mathcal{K}_1 \rightarrow \mathcal{K}_2)$ by performing 50 simulations starting from equilibrated polymers tied in a range of different knots. Some of the transition probabilities are reported in Table T\ref{table:tab}, both for free and confined polymers, and are compared with those reported by random passage~\cite{Flammini2004,Vazquez2007} (RP) and hooked juxtaposition~\cite{Burnier2007} (HJ) models (SI Appendix for full table). The transition rates towards simpler topologies outperform those of other TopoII-only models, in particular for more complex knots. For instance, to unknot a $7_1$ we predict the cascade $7_1 \rightarrow 5_1 \rightarrow 3_1 \rightarrow 0_1$ with probability $P(7_1 \rightarrow 0_1) = P(7_1 \rightarrow 5_1) P(5_1 \rightarrow 0_1) P(3_1 \rightarrow 0_1) = 0.98 $, which is about 12 times larger than the one predicted by RP models ($0.082$). This enhanced simplification with respect to RP and HJ models increases with knot complexity and with the degree of confinement. For instance, under the confinement chosen here, the RP model would predict a probability $P(7_1 \rightarrow 5_1 \rightarrow 3_1 \rightarrow 0_1) \simeq 0.002$ that is  about 300 times smaller than the one achieved by our synergistic model ($0.75$). 
	%This finding is intriguing, especially in light of the extreme packaging of the genome in the cell nucleus.

\begin{figure}[t!]
	\centering
	\includegraphics[width=0.45\textwidth]{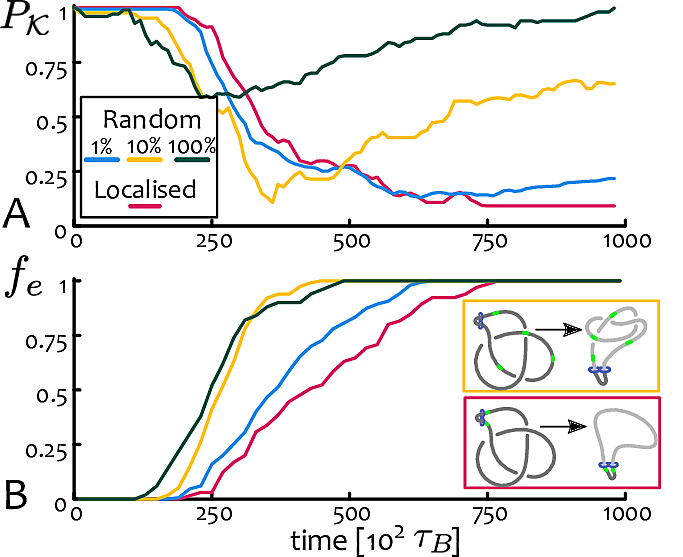}
	\vspace*{-0.2 cm}
	\caption{\textbf{Localised versus Random TopoII Under Confinement.} We perform simulations on a trefoil under confinement $R/\langle R_g \rangle = 1/3$ and measure (A) the knotting probability $P_\mathcal{K}$ and (B) the fraction of completed loops $f_e$ as a function of time. Our results show that while models of randomly bound TopoII can lead to substrate unknotting, they entail a return to equilibrium values of $P_{\mathcal{K}}$ once SMCs stop extruding. } 
	\vspace*{-0.5 cm}
	\label{fig:unkn_efficiency}
\end{figure}

\vspace*{-0.3 cm}
\subsection*{Randomly-Bound versus SMC-Localised TopoII}
While recent experimental data on SMC cohesin supports our hypothesis TopoII-SMC co-localisation~\cite{Uuskula-Reimand2016,Vian2018}, such evidence is poorer for condesin and bacterial SMC. Thus, we tested whether a model in which TopoII is dynamically and randomly associated with the polymer during SMC extrusion can still yield efficient topological simplification. We performed simulations of a confined trefoil in which a random fraction of contour length $\phi$ is allowed to undergo strand-crossing events and set the turnover time for TopoII-bound segments to be comparable to that taken to extrude one persistence length (SI Appendix).  	% $10 \tau_{B}$,
	
We discovered that the knotting probability $P_\mathcal{K}$ shows a non-monotonic behaviour as a function of time for all models of randomly associated TopoII (Fig.~\ref{fig:unkn_efficiency}A). By measuring the fraction of fully extruded loops $f_e$ we observed that the recovery of $P_\mathcal{K}$ at large times occurs after $f_e \simeq 1$. This is to be expected, since models with randomly associated TopoII must return to the equilibrium value for pure random passage events with $\phi$-dependent kinetics.
On the contrary, in our original  model where TopoII is only localised at the SMC, the successfully extruded polymer segments are no longer able to cross each-other and the topology is thus fixed at all future times. Thus, the 
recovery of $P_\mathcal{K}$ to its equilibrium values is neither expected nor observed. We thus argue that for randomly-bound TopoII a continuous flux of dynamically associated SMC is required in order to maintain a knotting probability below equilibrium.

\vspace*{-0.1 cm}

\vspace*{-0.2 cm}
\section*{Conclusions}
\vspace*{-0.1 cm}
In this work, we have provided numerical evidence for a new molecular mechanism that can efficiently maintain genomes free of entanglements. This is based on the combined action of SMC-driven extrusion and TopoII-mediated strand-crossing. The sliding of molecular slip-links along knotted or linked substrates naturally generates highly localised entanglements (Fig.~\ref{fig:panel1}) in turn catalysing their simplification through TopoII (Fig.~\ref{fig:smc-topo-model}), also under strong confinement (Fig.~\ref{fig:smc-topo-confined}). Importantly, the envisaged mechanism is universal, in that it works equally well on DNA or chromatin, closed plasmids or stably looped linear genomic regions such as TADs, interphase and mitosis and across all life forms that have evolved TopoII-like and SMC-like proteins. 

%Our findings provide a mechanistic explanation for existing evidence suggesting that the presence of both SMC condensin and TopoII, is required to ensure correct decatenation of sister chromatids~\cite{Baxter2012,Sen2016,Piskadlo2017} and that inactivation of SMC leads to an increase in entanglement complexity and sister chromatids intertwining~\cite{Piskadlo2017a}. 
%In our model, SMC-driven knot and link localisation is indeed an essential step to avoid the creation of more complicated knots and links which may occur under the extreme packing typically achieved in mitosis.

Our findings show that SMC proteins are indispensable to correctly decatenate sister chromatids, in agreement with experiments~\cite{Baxter2012,Sen2016,Piskadlo2017} and also shed light on recent findings reporting the accumulation of DNA damage in front of cohesin complexes~\cite{Vian2018}. We argue that the sliding motion of SMC entraps topological entanglements in turn increasing local stresses that may lead to double-strand DNA breaks. Our results thus provide compelling mechanistic evidence for an evolutionary optimal strategy whereby TopoII is actively recruited by SMC complexes~\cite{Uuskula-Reimand2016}. At the same time, we showed that randomly bound TopoIIs can still yield efficient topological simplification, if combined with dynamically associated SMCs (Fig.~\ref{fig:unkn_efficiency}).

Whilst we here assumed unidirectional SMC motion, we expect that similar physics should be at work for diffusing SMCs~\cite{Brackley2017prl} as the entropic competition between slip-links and knots may favour the former under some conditions~\cite{Zandi2003}; we aim to further explore this avenue in the future (SI Appendix).

We also argue that an analogous mechanism may take place during DNA replication, whereby the polymerising machinery effectively functions as a slip-link and localises entanglements. TopoII is known to act in front of the replication fork~\cite{Roca2008}, thus the very same synergistic mechanism for topological simplification proposed here may be at play in this context as well. 
It is also of interest to note that PCNA, the molecular clamp associated with a processive polymerase~\cite{Krishna1994}, recruits components of repair complexes, which would again be evolutionary advantageous to resolve entanglement-related DNA damage. All this reinforces the idea that the mechanism we propose may be universal.
%When two replication forks move along opposite directions effectively extrude a DNA or chromatin loop~\cite{Cook1999}. The polymerising machinery effectively functions as a SMC, driving the localisation of entanglements ahead of the growing loop. As TopoII is known to act in front of the replication fork~\cite{Roca2008}, the very same synergistic mechanism for topological simplification proposed here may be at play in this context as well. It is also of interest to note that PCNA, the molecular clamp associated with a processive polymerase~\cite{Krishna1994}, also recruits components of repair complexes, which would again be evolutionary advantageous to resolve entanglement-related DNA damage. All these considerations reinforce the idea that the mechanism we propose may be universal.

%provided that entropic considerations favour loop enlargement. This will be the case if SMC is loaded at particular sites, as this creates a positively biasing osmotic pressure~\cite{Brackley2017prl}, or when multiple slip-links are simultaneously present, as the entropy is generically maximised when some of the loops shrink to zero~\cite{Zandi2003}.

We finally speculate that the remarkable low knotting probability recently quantified in intracellular chromatin and its weak or absent scaling with the length of the substrate~\cite{Valdes2018} may be explained by our model as we find it to be remarkably insensitive to substrate length (SI Appendix). We hope that our work will ignite new experimental efforts to identify and further characterise novel synergistic mechanisms that may regulate genome topology. 

%For instance, we envisage that it may be tested {\it in vivo} by combining the action of TopoII and functional SMC~\cite{Kanke2016,Ganji2018} and \emph{in vivo} by quantifying chromatin knotting~\cite{Valdes2018} after knock-down of SMC. 

We conclude this work by speculating on an open question: if TopoII can co-operate with ATP-consuming SMCs to simplify genome topology, why does it require ATP to function (as shown in \emph{in vitro}~\cite{Rybenkov1997})? A possible explanation is that the synergy between passive TopoII and active SMC would still be insufficient to maintain a functionally viable genome in the cell nucleus. We hope that either, future models accounting for non-equilibrium 
TopoII or experiments exploring the synergy of TopoII and SMC, may shed light on this intriguing problem.

%% REOMBINATION WITH SMC -- COOL! XX TODO XX
%% Xer simplifies post-rep plasmids in absence of TopoIV. Yet it needs functional FtsK!!
%Wr also note that a similar mechanism may be at play during the XerCD-dif simplification in bacteria as this requires the processive action of a translocase (FtsK) to efficiently generate unlinked post-replicative chromosomes in \emph{E. Coli}. Notably, in absence of FtsK, the site-specific recombination by XerCD-dif produces more complex topological products~\cite{Shimokawa2013}. Inspired by the mechanisms uncovered in this work, we speculate that FtsK may be provide a way to  localise links which can then be efficiently simplified by the alignment of XerC/XerD recombination sites. 

%%CLIMBING ROPE ANALOGY
%We conclude this paper by drawing an old-standing analogy which inspired the first mechanistic model for the working of SMC proteins~\cite{Gruber2003}, i.e. that with traditional climbing gear. It is indeed amusing to notice that the reeling of long climbing ropes through carabiners is a well-established technique used by rock climbers to check for the presence of knots along their ropes. Indeed, this seems to be the most efficient way to discover and remove knots which, if undiscovered, could result fatal to the climbing team. Our work strongly suggests that this universal mechanism may also be employed across organisms to remove deleterious topological entanglements along genomes.   

\section{Methods}

\vspace*{-0.2  cm}
\subsection*{Chromatin/DNA Model}
We employ a well established bead-spring polymer model~\cite{Kremer1990} to describe chromatin and DNA~\cite{Rosa2008}. We account for excluded volume and chain uncrossability by using shifted and truncated Lennard-Jones interactions and finitely extensible springs to prevent thermally-activated strand-crossing events as discussed in the text (also SI Appendix). A publicly available code~\cite{Tubiana2018} is used to detect the shortest physically knotted arc within the substrate. 

\vspace*{-0.2 cm}
\subsection*{Integration Procedure}
Each bead in our simulation is evolved through the Langevin equation
$m_a \partial_{tt} \vec{r}_a  = - \nabla U_a - \gamma_a \partial_t \vec{r}_a + \sqrt{2k_BT\gamma_a}\vec{\eta}_a(t)$,where $m_a$ and $\gamma_a$ are the mass and the friction coefficient of bead $a$, and $\vec{\eta}_a$ is its stochastic noise vector satisfying the fluctuation-dissipation theorem. $U$ is the sum of the energy fields (SI Appendix). The simulations are performed in LAMMPS~\cite{Plimpton1995} with $m = \gamma = k_B = T = 1$ and using a standard velocity-Verlet algorithm. 
%We set the integration time step to be $\Delta t = 0.01\,\tau_{B}$, where $\tau_{B}$ is the Brownian time as mentioned previously. 

\section{Acknowledgements}
This work was supported by the ERC CoG 648050 THREEDCELLPHYSICS.
After the present paper was submitted for publication, we learnt of a similar model simultaneously developed by the group of A.~Stasiak and D.~Racko~\cite{Racko2018}. DMi and EO would also like to acknowledge the networking support by EUTOPIA (CA17139).

\bibliographystyle{pnas-new}
\bibliography{library}

\appendix
\newpage
\phantom{a}
\newpage
\section{SUPPLEMENTARY MATERIAL}

\section{Computational Details}

\subsection*{A Polymer Model for Chromatin and DNA substrates}
We model a polymer substrate, such as DNA or chromatin, as a chain of beads of size $\sigma$ connected by springs. This types of models are widely employed in the literature and have been shown to faithfully capture the physical behaviour of DNA and chromatin~\cite{Rosa2008,Sanborn2015,Goloborodko2015,Michieletto2016prx}. 
To ensure that the polymer substrates does not cross through itself, we impose that any two beads $(a,b)$ at distance $r$ are subject to a purely repulsive (WCA) potential
\begin{equation}\label{eq:LJ}
U_{WCA}^{ab}(r) = k_BT \left[ 4\left[\left(\frac{\sigma}{r}\right)^{12}-\left(\frac{\sigma}{r}\right)^{6}\right] + 1 \right] \, \text{if} \, r \le 2^{1/6}\sigma
\end{equation}
and 0 otherwise. Further, we impose that consecutive beads are connected by finitely extensible (FENE) springs modelled as
\begin{equation}\label{eq:Ufene}
U_{FENE}^{ab}(r) = - \frac{k_f R_0^2}{2}\ln\left[1-\left(\frac{r}{R_0}\right)^2\right] \, \text{if} \, r\le R_0 
\end{equation}
and $\infty$ otherwise. Here, $k_f = 30\epsilon/\sigma^2$ and $R_{0}=1.5\sigma$ are typical parameters employed to prevent spontaneous chain crossing~\cite{Kremer1990}. We account for DNA or chromatin stiffness by adding a potential controlling the angle formed by consecutive triplets of beads
\begin{equation}
U_{KP}^{ab} = \frac{k_BTl_p}{\sigma}\left[ 1- \frac{ \bm{t}_a \cdot \bm{t}_b }{ |\bm{t}_a | | \bm{t}_b| } \right],
\end{equation}
where $\bm{t}_a$ and $\bm{t}_b$ are the tangent vectors connecting bead $a$ to $a+1$ and $b$ to $b+1$ respectively; $l_p$ is the persistent length of the chain and by setting $l_p=20 \sigma=50$ nm we model an average DNA sequence~\cite{Calladine1997} while with $k_\theta= 3 \, k_B T$ we account for a more flexible polymer with $l_p = 3\sigma= 90$ nm such as a 30nm chromatin fibre~\cite{DekkerJob2002}.

\section*{A Model for Structural Maintenance of Chromosome Proteins}
The SMC proteins -- such as cohesins and condensins -- are a well-known and widely studied family of proteins~\cite{Hirano2016} that have now been identified as responsible for dynamic genomic loops in both inter- and meta-phase~\cite{Vian2018,Gibcus2018}. These proteins can be crudely viewed as physical slip-links~\cite{Michieletto2016softmatter} that embrace one, or two, double-stranded DNA and slide along DNA/chromatin~\cite{Brackley2017prl,Uhlmann2016} in turn stabilising the formation of dynamic loops~\cite{Fudenberg2016,Brackley2017prl} and halting at ``anchor'' points embodied by converging CTCF proteins~\cite{Rao2014,Vian2018}. 
In this work, we aim to mechanistically investigate the generic effect of SMC proteins on topological entanglements -- such as knots and links -- that may be present on DNA or chromatin in interphase and mitosis. To this end, we propose a generic model where SMC proteins loaded on the polymer are described as bonds connecting two non-consecutive beads along the chain. Importantly, and in marked contrast with recent models of loop extrusion~\cite{Fudenberg2016,Goloborodko2016,Sanborn2015}, here we account for the physical presence of a slip-link-like molecule joining two segments of chromosomes by forcing the maximum extension of the bond with a FENE potential so that it is energetically very unfavourable for a third bead to cross through the gap in between the joined segments. Again, this is done to prevent spontaneous events that would change the local topology of the substrate and that are not physically possible in real situations. 
It is worth noting that this detail had not been correctly accounted for in some of the existing models of loop extrusion~\cite{Fudenberg2016,Sanborn2015}. 
In other words, the SMC protein is modelled by including a potential 
\begin{equation}
U_{SMC}^{h_1 h_2}(r) = - \frac{k_f R_0^2}{2}\ln\left[1-\left(\frac{r(h_1,h_2)}{R_0}\right)^2\right] \, \text{if} \, r\le R_0 
\end{equation}
and $\infty$ otherwise, and where $h_1$ and $h_2$ are the instantaneous position of the two segments of chromosome bound by the SMC protein at time $t$ (or the SMC ``heads''). At rate $\kappa=0.01 \tau_B^{-1}$ ($\tau_B \equiv \sigma^2/D$ is the Brownian time of a bead, see below), we change the position of the heads via the following protocol:
\begin{equation}
\begin{cases}
&h_1(t+dt)=h_1(t)+1 \text{ and } \\
&h_2(t+dt)=h_2(t)-1 \, \text{ if } \, d(h_1,h_2) \leq 1.3 \sigma \\
&h_1(t+dt)=h_1(t) \text{ and } \\
&h_2(t+dt)=h_2(t) \, \text{ otherwise or if } h_1=h_2\, .
\end{cases}
\end{equation}
Thus, the SMC enlarges the loop formed by two monomers on average every $100$ Brownian times only if the distance between the next pair of beads is shorter than or equal to $1.3\sigma$ in 3D space. This choice ensures that no third bead can pass through the beads bonded by the SMC protein and it effectively slows down the speed of the complex from $v_{max}= 2 \sigma \kappa$ to about $v \simeq 0.1$ $v_{max}$. Unless otherwise stated, we will consider  $v_{max}= 0.02 \, \sigma/\tau_B$.
% or $v_{30nm}=0.125$ nm/s $=1.5$ kbp/min assuming a chromatin compaction of 100 bp/nm~\cite{Brackley2017prl}.
Yet, we stress that the speed of the extrusion (or diffusion) process does not affect the efficiency of the synergistic mechanism we uncover in this work. 

%Finally, we note that most of the work done here considers SMC complexes that are bound for infinite time but below we report some simulations with dynamically associated SMC (unloaded at rate $\kappa_{\rm off}$). In this case the relevant parameter is the processivity $p= v/ \kappa_{\rm off}$~\cite{Fudenberg2016}, i.e. the typical length covered by the complex before disassociating from the polymer.
 
%For simplicity we also focus on closed (circular) chains so that the knot (or link) cannot diffuse out of the polymer.  Yet, we stress that all our results are also valid for the case that a physical knot (or link) is tied on a long open DNA or chromatin segment interspersed with long-range loops mediated by bridge proteins~\cite{Brackley2013pnas} (Fig.~\ref{fig:panel1}E,F). 

\subsection*{Mapping to Real Units}
Given the size of a bead $\sigma$ and the energy scale $k_BT$ (at room temperature), we can derive the typical (Brownian) time taken for a bead to diffuse its own size as $\tau_B=\sigma^2/D=3 \pi \eta \sigma^3/k_B T$. Using the viscosity of the nucleoplasm $\eta \simeq 200 cP$~\cite{Baum2014} we obtain that our simulated Brownain time corresponds to $\tau_B=7 \mu$s for DNA and $\tau_B=12$ ms for a 30nm chromatin fibre. The initial state of our simulations is an equilibrated polymer conformation (without SMC proteins acting on it) that is obtained running $10^5$ $\tau_B$ steps, i.e. of the order of seconds for DNA and tens of minutes for chromatin. Production runs in which SMC proteins are loaded on the polymer also typically cover $10^5$ $\tau_B$ steps which we find is enough for complete knot localisation.

\subsection*{Topoisomerase Model}
In contrast to previous works which crudely model the action of TopoII as a uniform non-zero probability of strand-crossing events~\cite{Rosa2008,Goloborodko2016,Michieletto2014a}, here we assume that TopoII is locally recruited by the SMC protein and it is loaded on the outside of the loop subtended by the complex (see below for extensions of our model that relax this assumption). Thus, here only the two beads (about 60 nm) in front of the ones forming the SMC complex are allowed to undergo strand-crossing events. In simple terms, if $h_{1,2}(t)$ are the positions of the SMC heads then ($h_{1}(t) - 1$, $h_{1}(t)- 2$) and  ($h_{2}(t) + 1$, $h_{2}(t) + 2$) are the beads associated to TopoII. In practice, we set the interaction of these beads with all other beads as a soft repulsion
\begin{equation}
U_{Topo} = A \left[ 1+\cos{\left( \dfrac{\pi r}{r_c} \right)}\right]\, .
\end{equation}
To avoid numerical instabilities which may occur due to the dynamic update of the SMC heads, we tune $A$ so that it displays an increasing energy gradient, i.e. the furthest bead from the SMC complex is set to have $A=5 k_BT$ while the closer one $A=20 k_BT$. This ensures that when the position of the SMC is updated, it is unlikely for two beads interacting through the WCA potential to be overlapping.

\subsection*{Integration Procedure}
The total energy field experienced by bead $a$ is the sum of all the pairwise and triplet interactions involving all other beads, i.e.
\begin{align}
U_a = \sum_{b\neq a} &\left[ U_{WCA}^{ab} + \left( U_{FENE}^{ab} + U_{bend}^{ab} \right) \left( \delta_{b,a+1} + \delta_{b,a-1} \right) \right] + \notag \\
& + U^{ab}_{SMC}(t)\delta_{a,h_1}\delta_{b,h_2} \, ,
\end{align}
where the Kronecker deltas $\delta_{i,j}$ indicate that bond and angle potentials are restricted to consecutive beads along the polymer and that the SMC potential is acting on the beads corresponding to the SMC heads.
The time evolution of each bead in the system is thus governed by the following Langevin equation,
\begin{equation}
\label{eqn:EOM}
m_a\frac{d^2\vec{r}_a}{dt^2} = - \nabla U_a - \gamma_a \frac{d\vec{r}_a}{dt} + \sqrt{2k_BT\gamma_a}\vec{\eta}_a(t),
\end{equation}
where $m_a$ and $\gamma_a$ are the mass and the friction coefficient of bead $a$, and $\vec{\eta}_a$ is its stochastic noise vector obeying the following statistical averages:
\begin{equation}
\langle\bm{\eta}(t)\rangle = 0;\;\;\; \langle\eta_{a,\alpha}(t)\eta_{b,\beta}(t')\rangle = \delta_{ab}\delta_{\alpha\beta}\delta(t-t'),
\end{equation}
where the Latin indices represent particle indices and the Greek indices represent Cartesian components. The last term of Eq.~\eqref{eqn:EOM} represents the random collisions caused by the solvent particles and, for simplicity, we assume all beads have the same mass and friction coefficient (i.e. $m_a = m$ and $\gamma_a = \gamma $) and finally set $m = \gamma = k_B = T = 1$.  Equation \eqref{eqn:EOM} is integrated using a standard velocity-Verlet algorithm, which is performed using the Large-scale Atomic/Molecular Massively Parallel Simulator (LAMMPS)~\cite{Plimpton1995}. For the simulation to be efficient yet numerically stable, we set the integration time step to be (unless otherwise stated) $\Delta t = 0.01\,\tau_{B}$, where $\tau_{B}$ is the Brownian time mentioned previously.

\section{Circular Polymers Mimic Plasmids or Stably Looped Genomic Regions}
Our choice to consider circular polymers as substrate for the synergistic action of SMC and TopoII is motivated by the following arguments. First, circular genomes exist in nature, for instance plasmids in bacteria and mini-rings in the kinetoplast DNA~\cite{Chen1995a} (our polymers would correspond to a DNA molecule $L = N \sigma= 1.25 \, \mu m \simeq 4 $ kbp long if taking $\sigma= 2.5 $ nm). Second, although eukaryotic genomes are not topologically closed such as bacterial plasmids, they are transiently looped by bridge proteins~\cite{Alberts2014}. In particular, the size of the polymers considered here would map to $L=500$ kbp if coarse-graining a chromatin fibre with thickness $\sigma = 10 $ nm $= 1$ kbp (tightly packed chromatin) or $L=100 kbp$ if considering $\sigma = 10 $ nm $= 200$ bp (loosely packed chromatin). The length of so-called ``Topologically Associated Domains'' (TADs) in humans ranges from 40kbp to 3Mbp with a median of 185 kbp~\cite{Rao2014}; thus the polymers considered here would represent typical TADs. Increasing evidence suggest that TADs are stably looped by CTCF complexes~\cite{Rao2014,Nora2017,Rao2017} and therefore any non-trivial topological state (a knot or a link) assumed by TADs would be topologically trapped as long as the TAD itself is looped. 

For these reasons we argue that our choice of polymer size and global topology (that of a ring) correctly capture the length-scales and topological problem faced in vivo by bacterial and eukaryotic cells.

\begin{figure}[t]
\includegraphics[width=0.45\textwidth]{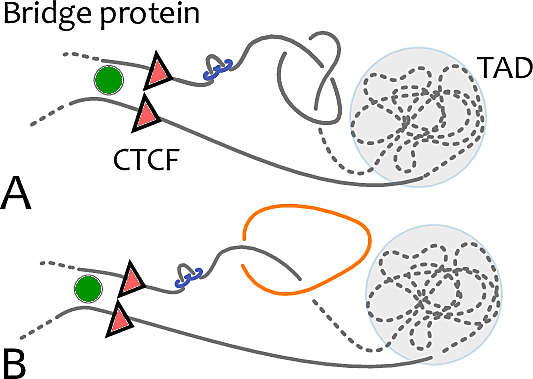}
\caption{Considering a circular polymer as a substrate not only mimics bacterial plasmids and DNA mini-rings but also stably looped genomic regions in eukaryotes. These regions may be looped by chromatin-binding proteins such as transcription factors (e.g., HMGB2~\cite{Zirkel2017}) and Polymerases, but also by CTCF~\cite{Rao2017} in the case of TADs.  }
\label{fig:tadloop}
\end{figure}

\section{Knot Inversion via SMC Extrusion}
\begin{figure}[t]
\includegraphics[width=0.45\textwidth]{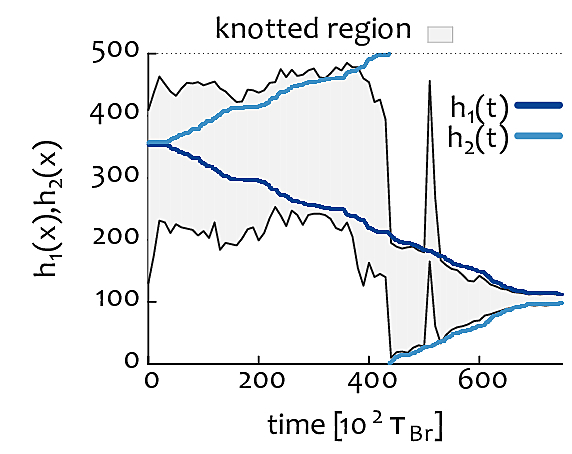}
\caption{The figure shows a kymograph of the shortest knotted segment (grey shade and black lines) together with the instantaneous location of the SMC heads (shades of blue). In particular, this kymograph shows an event of knot inversion by loading a SMC protein within the shortest knotted arc at time 0.} 
%Notice that the abrupt change in in the position of the knot (spike) is likely an artefact of the localisation algorithm. Nonetheless, these events are very rare and instantaneous so that they can be easily removed a posteriori if required.}
\label{fig:knotinversion}
\end{figure}

In this section we report a localisation events that starts from a SMC loading occurring within the shortest knotted arc. Such an event is not unlikely, since (i) the knot is originally delocalised and thus occupies a non-negligible contour length of the polymer and (ii) the loading of a SMC is a local event that cannot measure non-local topology. We inspected out simulations and in Fig.~\ref{fig:knotinversion} we report a kymograph of one such an event. As one can notice, the boundaries of the knot are first inflated and then, because of globally closed topology of the underlying polymer, wrapped around and collapsed.
This event thus lead to full knot localisation as the case in which the SMC is loaded outside the shortest knotted arc. 

\section{Localisation Efficiency as a function of Number of SMC}
\begin{figure*}[t!]
	\centering
	\includegraphics[width=0.8\textwidth]{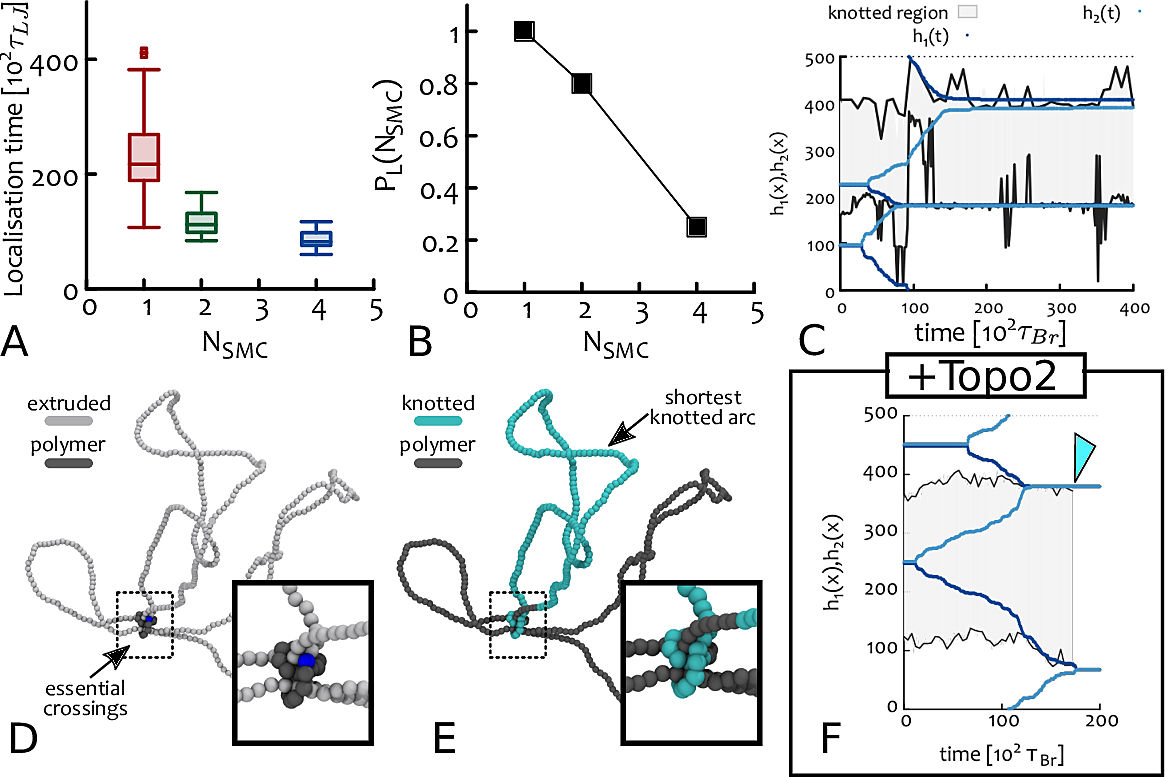}
	\caption{(A) Box plots showing the distribution of localisation times on a polymer substrate $N=500$ beads long and with SMC moving at rate $\kappa=\tau_{Br}^{-1}$. [Here a knot is considered localised if its shortest arc spans less than 50 beads]. (B) Probability to find a localised knot at large times as a function of the number of SMC proteins simultaneously loaded on the substrate. (C) Kymograph of a simulation for a polymer tied in a trefoil knot and with 2 SMC loaded at time $t=0$. The shortest arc that can be defined knotted (after suitable closure~\cite{Tubiana2011prl}) is shaded in grey. The 4 SMC heads are shown in shades of blue. One of the two SMC is loaded within the knotted region at time $t=0$, whereas the other is outside it. Eventually, heads belonging to different SMC meet and stall, thereby stabilising a delocalised knot. (D) Snapshot from the same simulation used to generate (C) and showing extruded segments in light grey and the polymer backbone in dark grey. One can see that essential crossings are localised (inset). (E) Snapshot highlighting the shortest arc that can be defined knotted in blue, while the rest of the polymer is showed in dark grey. These snapshots shows that even if essential crossings are localised, the knot itself need not be. This conformation has been recently found in simulations of translocation of knotted DNA~\cite{Suma2017}).  See Supplementary Movie M6. (F) Including Topo2 localised in front of SMC leads to the unknotting of a knot whose essential crossings are localised. The arrowhead in \textbf{F} indicates the unknotting event. }
	\label{fig:loc_vs_nlefs}
\end{figure*}

It is reasonable to ask whether the synergistic effect we uncover in this work may be made more efficient by considering multiple SMC extruding loops on the same substrate. To answer this question we perform simulations in which we simultaneously load 1, 2 and 4 SMC complexes at a random position along a polymer which is tied in a trefoil knot. The interaction between SMC heads is here considered mutually exclusive, i.e. if two SMC heads are found on consecutive beads and moving in opposite directions they remain still as cannot overlap on the same bead. In these simulations we discover two seemingly counter-intuitive effects: 
\begin{enumerate}
	\item the knots which become localised do so in shorter time when multiple SMC are loaded  (Fig.S~\ref{fig:loc_vs_nlefs}A); 	
	\item the probability to find a localised knot (here practically defined as one made by less than 50 beads) at large times decreases with the number of SMC (Fig.S~\ref{fig:loc_vs_nlefs}B);
\end{enumerate}
The former finding can be readily explained by the fact that multiple SMC can extrude more contour length on the same unit of time. Yet, the decrease in localisation probability is more puzzling.  

By close inspection of the simulation trajectories we discover that this reduction in localisation probability is due to situations in which two or more SMC proteins are simultaneously loaded within and outside a knotted region. These situations may lead to trapped conformations that stabilise a delocalised knotted state (see kymograph and snapshots Fig.~\ref{fig:loc_vs_nlefs}).  
On the contrary, a single SMC, even if loaded within a knotted region can turn the knot ``inside-out'' and ultimately generate a fully localised knot.

\section{Unknotting is Favoured Under Confinement}
As mentioned in the main text, we find that the proposed synergistic mechanism between SMC and TopoII can simplify knots even under strong confinement. Because of this, we argue that this pathway may be at work in vivo. Here, we further characterise this finding by quantifying the rate of knot localisation as a function of confinement. To this end, we perform different sets of 40 independent simulations in which a trefoil knot tied along a $N=300$ beads polymer is confined within a sphere of varying radius $R_c$ and subject to the action of a single SMC. We consider a range of values for $R_c$ ranging from tight confinement $R_c=10 \sigma \simeq \langle R_g \rangle/3$ to $R_c=50\sigma > \langle R_g \rangle$, where $\langle R_g \rangle$ is the typical size of the polymer in equilibrium in good solvent and under no confinement. 

\begin{figure}[t]
	\centering
	\includegraphics[width=0.45\textwidth]{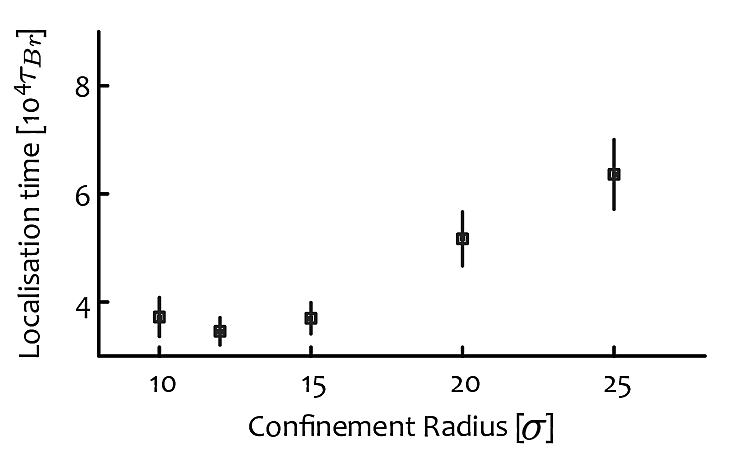}
	\caption{Average localisation time of a trefoil knot tied along a polymer $N=300$ beads long, under confinement within a sphere and subject to the action of a single SMC moving at rate $\kappa=0.1 \tau_{Br}^{-1}$. The error bars represent the standard error of the mean.}
	\label{fig:unkn_vs_conf}
\end{figure}

Remarkably, we discover that the typical localisation time (here practically defined as the first time at which the shortest knotted arc spans less than 50 beads) is shorter the stronger is the confinement (see Fig.~\ref{fig:unkn_vs_conf}). We argue that this puzzling finding can be explained by the following argument: the entropic penalty associated with the formation of a loop of length $l$ is $S/k_B \sim c \log{l}$ where $c$ is the exponent determining the decay of the contact probability $P_c(l) \sim l^{-c}$. For a crumpled polymer, i.e. the conformation assumed under confinement, the contact exponent $c = \nu d = 1$, whereas for a swollen coil in good solvent (self-avoiding), $c \simeq 2.1$~\cite{Gennes1979}. For this reason, the entropic penalty grows more steeply for a swollen coil than for a crumpled globule. In turn, this implies that the loop extruding action of the SMC protein is entropically favoured (or less hindered) under confinement, in qualitative agreement with our findings (see Fig.~\ref{fig:unkn_vs_conf}).

\section{Synergistic Unknotting is Insensitive to Substrate Length}

In this section we provide a more quantitative, albeit not definitive, examination of efficiency of the proposed synergistic simplification as a function of the length of the substrate. Because of the largely fluctuating 3D conformations assumed by long polymers, the random passage and hooked juxtaposition models are known to be sensitive on this parameter~\cite{Vologodskii2016}. To compare these models with the one proposed here, we perform 4 sets of 50 independent simulations starting from an equilibrated (and unconfined) polymer tied as a $7_1$ knot with varying length $M$, ranging from $300$ to $2000$. If one takes $\sigma=2.5$ nm as the diameter of DNA, then this range compares to $750-5000$ bp. In Table T~\ref{table:tabsi} we report the values of the transitions $k_1 \rightarrow k_2$ observed for these different lengths of the substrate. 

\begin{table}[t!]
	\centering
	\begin{tabular}{c | c | c | c | c} 
		\toprule
		& $M=300$ & $M=500$ &  $M=1000$ & $M=2000$ \\ 
		\hline 
		\hline
		$7_1 \rightarrow K$ & $0$ & $0.02$ & $0$ & $0.1$ \\	
		$7_1 \rightarrow 5_1$ & $1$ & $0.98$ & $1$ & $0.9$ \\	
		\hline
		$5_1 \rightarrow K$ & $0$ & $0$ & $0$ & $0.03$ \\	
		$5_1 \rightarrow 3_1$ & $1$ & $1$ &  $0.97$ & $0.97$ \\	
		$5_1 \rightarrow 0_1$ & $0$ & $0$ &  $0.03$ & $0$ \\	
		\hline
		$3_1 \rightarrow K$ & $0$ & $0$ & $0$ & $0.03$ \\	
		$3_1 \rightarrow 0_1$ & $1$ & $1$ &  $1$ & $0.97$ \\	
	\end{tabular}
	\caption{Comparison of unknotting pathways for a $7_1$ knot tied along a polymer of length $M$ in unconfined conditions and subject to the synergistic unknotting. To obtain this table, 50 independent simulations were initialised from a $7_1$ state and the transition to other knot types recorded. $k_1 \rightarrow K$ denotes a transition to any knot type with minimal crossing number larger than $k_1$. 
	}
	%One can notice that the efficiency of knot simplification is only mildly reduced by increasing the polymer length. }
	\vspace*{-0.4 cm}
	\label{table:tabsi}
\end{table}

As one can notice, we find that for $M=300$ the $7_1$ is taken to the unknot through $5_1$ and $3_1$ with probability 1 and this probability is only mildly, if at all, affected for longer substrates. We highlight that this finding is likely to due to the fact that TopoII strand-crossing occurs more likely in pre-localised topological entanglements, thus strongly biasing their simplification over the increase in complexity. 

In other words, while the time to localise a knot increases on longer substrates, the simplification cascade towards the unknot is virtually unaffected. In light of this insensitivity, we reason that if this mechanism is at work in vivo, then the knot probability in intracellular chromatin should only weakly depend on the length of the fibre under consideration. Intriguingly, this observation is consistent with very recent experimental findings on the knotting of chromatin fires in vivo~\cite{Valdes2018}. 

\section{Loading and Unloading SMC Complexes} 

SMC proteins, cohesin and condensin, have a finite residency time on chromatin~\cite{Tedeschi2013,Hansen2017}. For cohesin, this is typically of the order of $\tau=20$ minutes~\cite{Busslinger2017}. On the contrary, the model considered up to now assumed that SMC would never disassociate from the substrate. This assumption is well justified only in the regime in which SMC proteins cover a length at least equal to the polymer size $L$ before disassociating. We now show that this condition is met for the cases considered in this paper: considering a thickness of 10nm (typical for loosely packed chromatin) our polymer made of $500$ beads describes a $L=100-500$ kb segment (by coarse graining either 200 bp or 1 kbp into 10 nm). Typical TAD sizes are in the range $40$ kb - $3$ Mbp, with a median of $185$ kb~\cite{Rao2017}. Thus our polymer represents a typical TAD in the scenario of loosest compaction (200 bp = 10 nm), or a large one in the case of tight compaction (1 kbp = 10 nm). In practice, we argue that in vivo chromatin compaction is heterogeneous, and that our polymer is well representative of a typical TAD in vivo. 

With these numbers in mind, one should now consider typical extrusion (or diffusion) speeds of cohesin and condensin complexes in order to predict whether a SMC protein would be able to extrude a loop $\simeq L$ before disassociating at time $\tau$. A useful parameter to bear in mind in this context is the processivity~\cite{Fudenberg2016} $p= v \tau$ which captures the typical distance covered by a unidirectional SMC within time $\tau$ (for a diffusing SMC this parameter is equivalent to $p=\sqrt{D \tau}$).

\subsection*{SMC can actively or diffusively extrude loops comparable to large TADs before disassociating}

We recall that recent experiments in vitro on condensin~\cite{Ganji2018} measured an extrusion speed of at least $0.6$ kb/s whereas indirect measurement using HiC in vivo obtained $0.2$ kb/s~\cite{Gibcus2018} for eukaryotic and $0.9$ kb/s~\cite{Wang2017} for bacterial condensin, respectively. It is thus ready to compute the range of distances travelled by condensin before disassociating (using $\tau= 20$ min): 240 kbp~\cite{Gibcus2018}, 720 kbp~\cite{Ganji2018} and 1Mbp~\cite{Wang2017}. These numbers are systematically larger than the size of chromatin fibre considered in this work which correspond to typical TADs in vivo. We thus argue that the assumption of permanently loaded SMC is a good approximation for typical TADs in vivo.

In the case of cohesin in interphase, in vitro experiments could not find unidirectional motion but measured an apparent diffusion constant of $1.72 \mu m^2/s$~\cite{Davidson2016}, $3.8 \mu m^2/s$~\cite{Stigler2016} and $0.25 \mu m^2/s$~\cite{Kanke2016}. A lower and upper bound of diffused lengths within $\tau=20$ min and chromatin compaction $C = 0.1$ kb/nm (1kbp=10nm) are $p\simeq 1.7 - 6.7$ Mbp. [The minimum mobility to span 500 kbp in 20 minutes via diffusion would be $D=0.02 \mu m^2/s$ at this chromatin compaction]. Whereas for the loosest chromatin fibre $C = 0.02$ kb/nm (200 bp = 10 nm) the range of distance covered is $p\simeq 0.35 - 1.35$ Mbp. 

As one can notice, diffusion of a cohesin over a chromatinised substrate can effectively span larger TADs than unidirectional condensin stepping (or hopping) on naked DNA~\cite{Brackley2017prl}. In particular, we find that for both diffusing and actively extruding SMC, one can safely think SMC proteins to be permanently loaded over substrates with length of typical TADs in vivo (about 200 kbp). It is also intriguing to notice that ``stripes'' in HiC maps which are linked to cohesin are most abundant in TADs within this range of lengths~\cite{Vian2018}. 

This approximation breaks down at the length scale of very large TADs, as neither extruding nor diffusing cohesins can \emph{systematically} cover length-scales of more than about $L \sim 1$ Mbp before disassociating. For this reason, in the next section we perform additional simulations of dynamically loaded SMC in order to study the limits of the proposed topological simplification mechanism and show that it is still valid in a stochastic sense rather than a systematic one.

\begin{figure*}[t]
\includegraphics[width=0.8\textwidth]{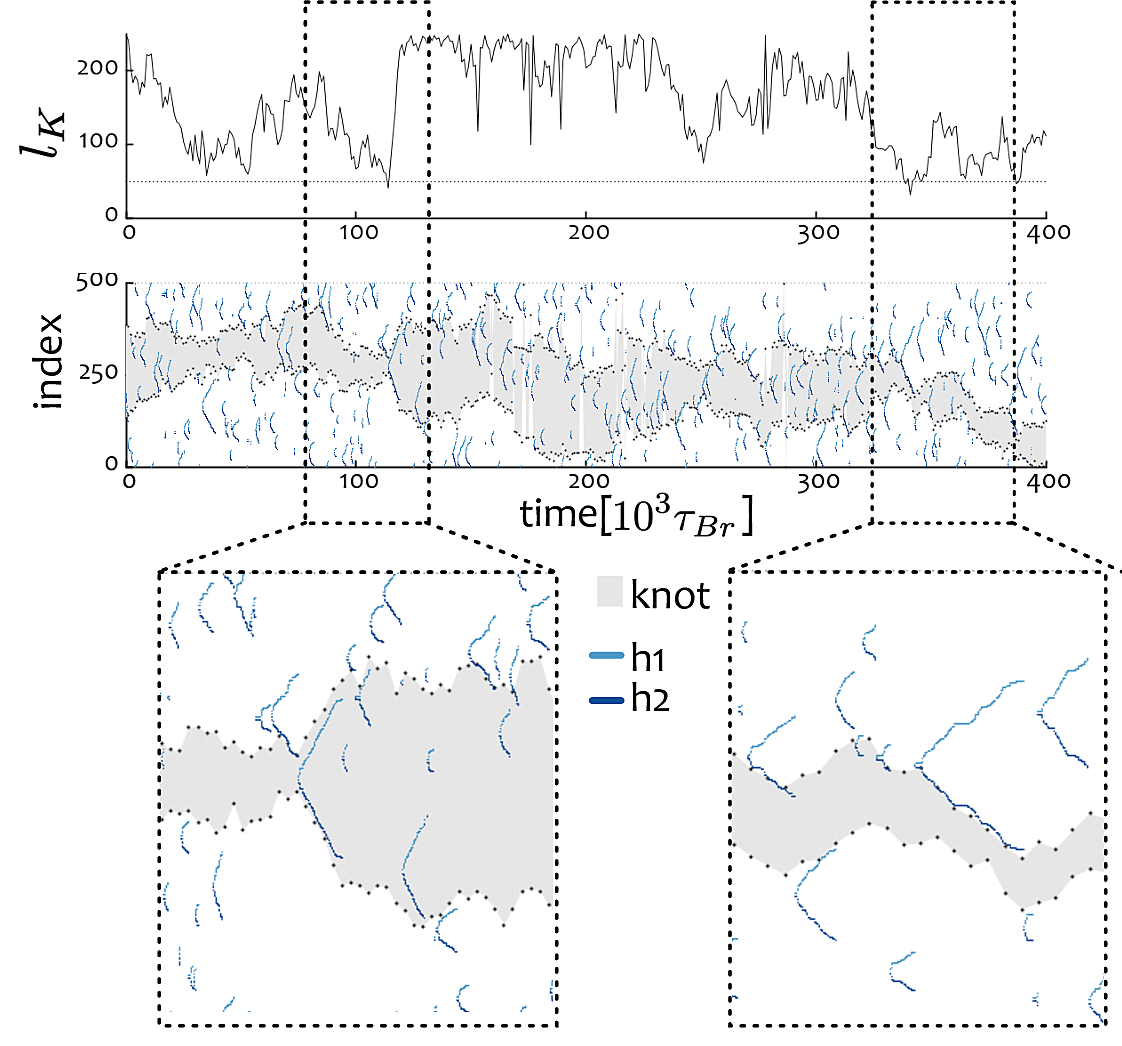}
\caption{Randomly loaded SMC proteins with processivity shorter than the polymer size (here we set $p_{max}=100 \sigma$ and effective processivity $p \simeq 25 \sigma$ on a polymer of length $L=500 \sigma$, see text for more details). The mid panel shows a kymograph of the knotted region (grey) and the position of SMC heads (blue). The top panel shows the instantaneous length of shortest knotted arc $l_K$ (dashed line marks $l=50$ beads). Bottom panels are zoom ins highlighting two different events: on the left a SMC is loaded within the minimum knotted arc and expands it, on the right several SMC are loaded on the flanks of the knot and reduce its size. }%Maximum speed is set to $v_{max} = 0.2 \sigma/\tau_B$. }
\label{fig:processiveSMC1}
\end{figure*}

\subsection*{Polymer Statistics and Unknotting via Processive SMC}
\begin{figure}[t]
\centering
	\includegraphics[width=0.4\textwidth]{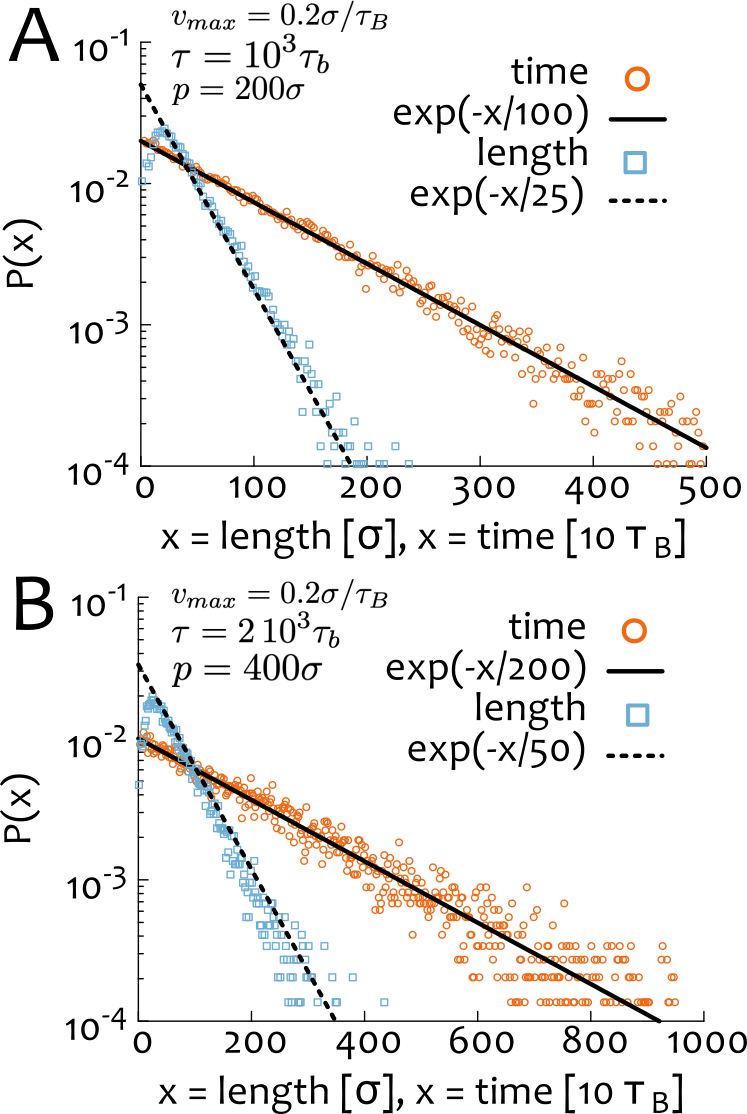}
	\caption{Distribution of SMC residency times (orange circles) follows an exponential decay with typical time (A) $\tau=10^3 \tau_B$ and (B) $\tau=2 \, 10^3 \tau_B$. The maximum speed is here set to $v_{max} = 0.2 \sigma/\tau_{B}$ so that one expects a maximum processivity (A) $p= v \tau = 200 \sigma$ and (B) $p= v \tau = 400 \sigma$. The distribution of lengths covered by the SMC (blue squares) is instead captured by an exponential with effective processivity about (A) $p=25 \sigma$ and (B) $p=50 \sigma$ (corresponding to an effective velocity $v= 0.125 v_{max}$). The peak at short lengths is due to the chosen persistence lengths of about 20 beads.}
	\label{fig:distr_length_proc}
\end{figure}

In this section we discuss the results from several sets of simulations in which SMC proteins are dynamically loaded and unloaded at a certain rate $\kappa=1/\tau$. Every time a SMC disassociates, we load a new one so that there is always one SMC bound at any one time (see Fig.~\ref{fig:processiveSMC1} for an example of a kymograph). The disassociation time $\tau$ gives an upper bound on the length covered by an SMC with maximum speed $v_{max}$, which we here choose to be $v_{max}=0.2 \sigma/\tau_B$, i.e. an update on the position of both SMC heads is performed every 10 Brownian times. In this model, the \emph{time} spent by any one SMC on the polymer follows a Poissonian statistics with mean $\tau$, yet because of our conditional rule on the update move, the effective speed is $v < v_{max}$. This implies that the effective processivity is also shorter than the maximum one, i.e. $p = v \tau < p_{max}= v_{max} \tau$.

By measuring the distance covered by SMC at the unloading event, we recover the real distribution of lengths spanned alongside the real residency time of the SMCs. These are reported in Fig.~\ref{fig:distr_length_proc} for two choices of $\tau$.  We observe that the effective speed $v \simeq 0.125 v_{max}$ so that the real SMC processivity $p$ is about 8 times shorter than the one set externally, i.e. $p = p_{max}/8$.  

As shown in Fig.~\ref{fig:distr_length_proc}, the statistics of unloading times and lengths correctly follows a Poissonian process; thus, the distribution of covered lengths is 
\begin{equation}
P(l) = \dfrac{1}{p}e^{-l/p}
\end{equation}
where $p$ is the processivity. The probability to observe an event with residency time longer than $\tau^\prime$ and hence length covered larger than $l^\prime$ is
\begin{equation}
p_{>l^\prime} = \int_{l^\prime}^{\infty} P(l) dl = 1 - \int_0^{l^\prime} P(l) dl
\end{equation}
and the typical number of events required to observe one such an event is simply $1/p_{>l^\prime} = e^{l^\prime/p}$. 

\begin{figure}[t]
\includegraphics[width=0.45\textwidth]{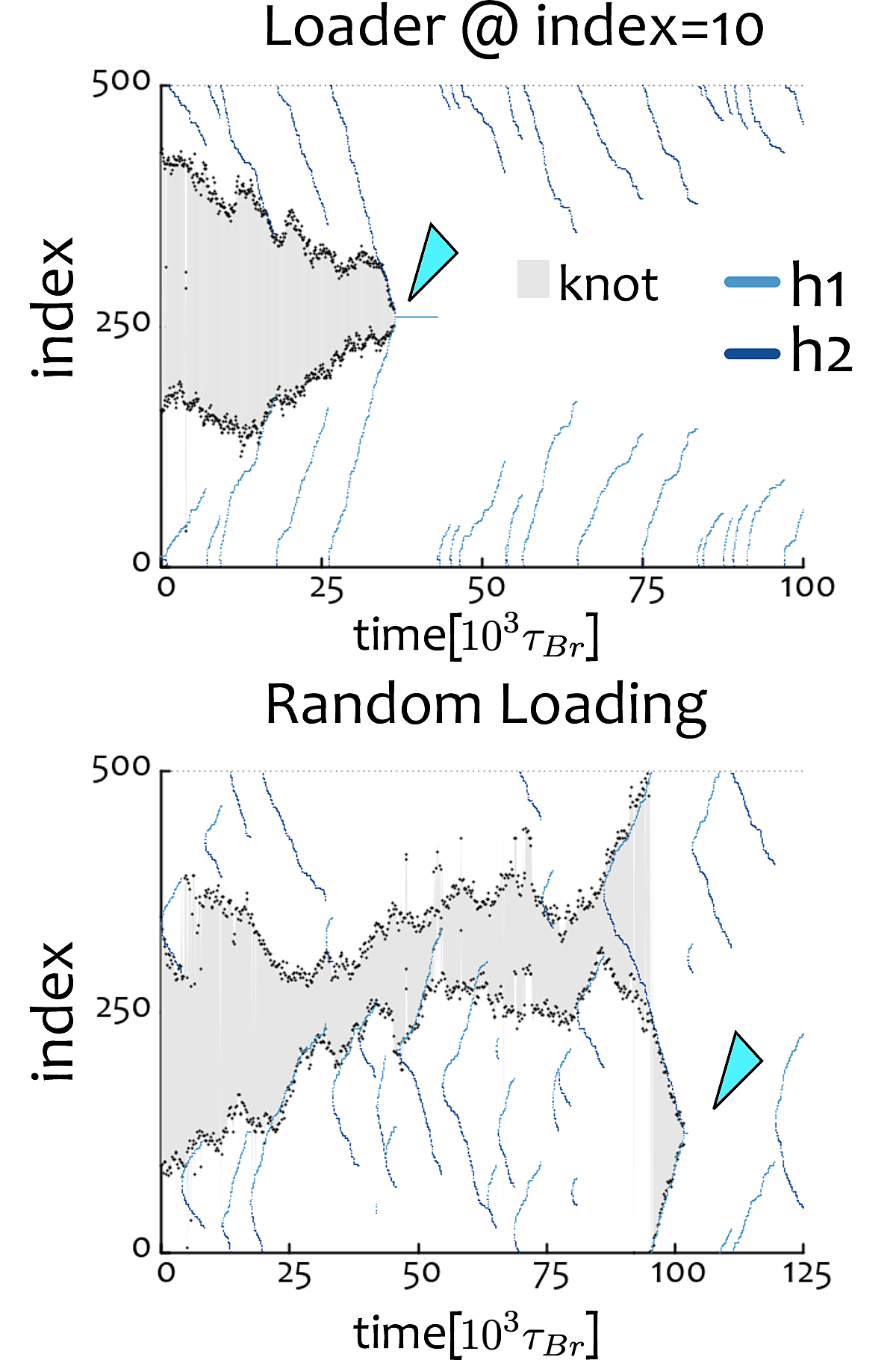}
\caption{SMC proteins with effective processivity $p = 125 \sigma < L = 500 \sigma$ achieve stochastic efficient unknotting of a trefoil. The system needs to sample about $n_e= 1/p_{>L} = e^{L/p} \simeq 50 $ events before observing one spanning the full contour. The figure shows a kymograph of the knotted region (grey) and the position of SMC heads (blue). The top panel shows the case of fixed loading (at bead 10) while the bottom panel shows the case with random loading. Arrowheads point to the unknotting event.  }%Maximum speed is set to $v_{max} = 0.2 \sigma/\tau_B$.}
\label{fig:processiveSMC2}
\end{figure}

For instance, for $p=30$ and $l^{\prime}=500$ one is required to sample $n_e \sim e^{17} \sim 25 \, 10^6$ events in order to observe one with processivity longer than 500 beads.  For $p=125$ this sampling number goes down to $n_e \simeq 50$, and indeed this is roughly what we find to be enough in order to untie a knotted polymer via SMC with  effective processivity $p=125<L=500$ (see Fig.~\ref{fig:processiveSMC2}). 

Accordingly, to localise and then untie a knot on a $2$ Mbp TAD through a SMC with speed $0.6$ kbp/s and residency time $\tau=20$ min one needs to sample on average 16 events, which may be roughly compatible with (if not underestimating) the number of cohesins loaded on a single TAD through interphase. We thus argue that while the approximation of infinite residency time is not kinetically accurate when $p<L$, the unknotting mechanism is still valid at times large enough to sample $e^{L/p}$ SMC loading/unloading events.   

It is finally worth stressing that loading multiple (non-nested) SMC will linearly accelerate this process further.

\section*{Randomly-Bound versus SMC-Localised TopoII}
To study the case in which TopoII is randomly bound on the substrate (Fig.~4 of main text) we performed at least 100 simulations for each value of TopoII density $\phi=1,10,100\%$ and compared the knotting probability obtained by averaging over from 100 simulations done with a SMC-localised TopoII. 
 
To do these simulations we started with a confined trefoil knot with $N=500$ beads and either (i) loaded one extruding SMC at random and placed beads with soft interactions in front of SMC or (ii) loaded one extruding SMC at random and placed $\phi N$ TopoII beads which we set as having soft interactions with the others. 

As done in previous cases the SMC heads were updated every $10 \, \tau_{B}$ steps, meaning a maximum speed $v_{max} = 0.2 \sigma/\tau_B$ or a real speed of about $v=v_{max}/8$ (see above). In the case of dynamical and randomly bound TopoII, the soft-interacting beads were set to be dynamically replaced along the substrate on average every $\tau_T=10^3 \tau_B$ meaning that the SMC complex can effectively cover a length $v_{max}\tau_T/8 \simeq 20 \sigma$. For the case with $\phi=1$ we simply set all the beads to be softly repelling all others, thus allowing bead-bead overlaps.  

\section{Diffusive SMCs can Undo Knots by Localising Essential Crossings}
To model diffusing SMCs, we update the position of the two heads independently, and move them either forward or backward with probability $1/2$. As before, the actual update is still conditional to the new Euclidean distance being shorter than the FENE bond.  In order to speed up the simulations, we consider a shorter substrate ($L=100$ beads $=100$ kb) and perform update moves every Brownian time. This is equivalent to a maximum mobility of $1 \sigma^2/\tau_{Br}= 200 kbp^2/s = 0.02 \mu m^2/s$ (using $\tau_{Br}=0.5 ms$ for a 10nm fibre), which is still smaller than that of cohesin in vitro~\cite{Kanke2016}. We further recall that due to our conditional updating rule on the position of SMC heads, the actual simulated diffusion is even slower than this value.
% but it is nonetheless compatible with the diffusion of a $500$ kb segment (and amply enough for a 100 kb one) in 20 minutes of residency time (see above).    

\begin{figure}[h]
	\includegraphics[width=0.5\textwidth]{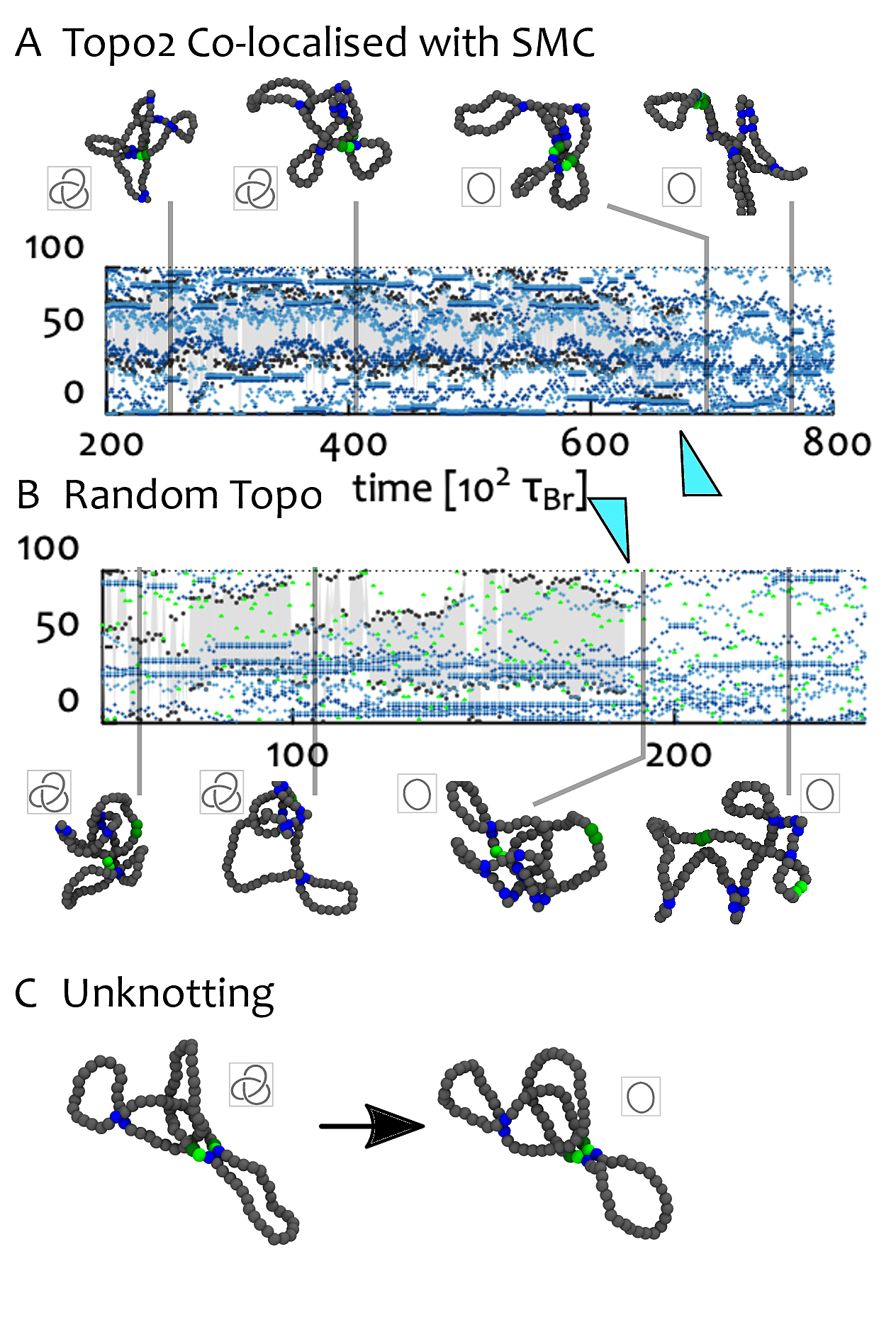}
	\caption{We perform simulations of diffusive SMC proteins on a 100 beads long polymer (tied in a trefoil knot) with co-localised (\textbf{A}) and random (\textbf{B}) action of Topo2. Both sets of simulations can yield unknotted substrates. Since we do not observe the localisation of the knotted arc, we argue that this unknotting pathway proceeds through the localisation of essential crossings, as more clearly illustrated in Fig.~\ref{fig:loc_vs_nlefs}E. Thus, we require $N_L>1$ in order to unknot polymers. In \textbf{C} we show an unknotting event with $N_L=2$. In the snapshots, blue beads mark SMC heads while green beads are TopoII-bound segments.}
	\label{fig:dLEF_1}
\end{figure}

We perform simulations starting from a trefoil knot, randomly load $N_L=5$ SMC proteins and place 2 soft repulsive beads either (i) located in front of one of the $N_L$ SMC (picked at random at every update timestep) or (ii) located randomly along the contour. In practice, to avoid numerical instabilities, case (ii) is modelled by setting a random pair of consecutive beads as ``Topo2-active'', i.e. subject to soft repulsive potential with the other beads with maximum energy barrier $\epsilon=4 k_BT$ and before returning them to the Lennard-Jones potential, they are transiently set to a ``Topo2-removing'' state in which they still interact via a soft potential but with a larger repulsive barrier ($\epsilon=20 k_BT$).

We discover that both sets of simulations yield to unknotting (see Fig.~\ref{fig:dLEF_1}) and that this process is not anticipated by the localisation of the knotted arc, but by through the localisation of the essential crossings as also seen for the case of extruding SMCs (Fig.~\ref{fig:loc_vs_nlefs}C-E).  In Fig.~\ref{fig:dLEF_1}C we show two consecutive snapshots in which the trefoil is being untied from a substrate with $N_L=2$ diffusive SMCs.

\vspace*{0.5 cm}
\section{Table of Transitions in Knot Space}
In Table ~\ref{table:tabSI} we report a transition rates for all the topologies studied in this work. 
We recall that these transitions are calculated by performing at least 50 simulations initialised with a given topology. Every time the knot changes topology we record the event and finally compute the probability to end up in another knotted state. From the table it is evident that torus knots follow ``cascades'' whereas twist knots ($7_2$ and $5_2$) have a non negligible probability to be unknotted in one step. Rates from random passage and hooked juxtaposition models are obtained from other works as shown in the table. These models extract transition rates on freely diffusing and flexible polymers. To faithfully compare transition rates in the confined case we thus performed simulations of the random passage model under the same confinement conditions as the ones for the synergistic case. See main text for detailed discussion.

\begin{table*}[t!]
	\centering
		\setlength\tabcolsep{8 pt} % default value: 6pt
	\begin{tabular}{c | c c | c c | c | c } \toprule
		& \multicolumn{2}{c}{Synergistic (this work)} & \multicolumn{2}{c}{RP } &  HJ & RP \\ 
		% \hline
		& Free & Confined  & Free Ref.~\cite{Flammini2004}  & Free Ref.~\cite{Vazquez2007}  & Free Ref.~\cite{Burnier2007} & Confined (this work)\\ 
		\hline
		%% NOW OK FREE AND CONFINED
		$7_2 \rightarrow K$ & $0$ & $0.02$ & -- & $0.5$ & -- & -- \\
		$7_2 \rightarrow 5_2$ & $0.31$ & $0.43$ & -- & $0.25$ & -- & -- \\
		$7_2 \rightarrow 5_1$ & $0$ & $0.02$ & -- & $0.005$ & -- & -- \\
		$7_2 \rightarrow 0_1$ & $0.69$ & $0.52$ & -- & $0.24$ & -- & -- \\
		\hline
		%%NOW OK FREE AND CONFINED
		$7_1 \rightarrow K$ & $0.02$ & $0.06$ & -- & $0.66$ & -- & $0.98$ \\	
		$7_1 \rightarrow 5_1$ & $0.98$ & $0.92$ & -- & $0.34$ & -- & $0.02$\\	
		$7_1 \rightarrow 3_1$ & $0$ & $0.02$ & -- & $0$ & -- & 0 \\	
		\hline
		%% OK FREE AND CONFINED
		$6_1 \rightarrow K$ & $0$ & $0.04$ & $0.15$ & $0.3$  & -- & -- \\
		$6_1 \rightarrow 4_1$ & $0.39$ & $0.33$ & $0.53$ & $0.44$ & -- & -- \\
		$6_1 \rightarrow 0_1$ & $0.61$ & $0.63$ & $0.32$ & $0.26$ & -- & --\\ 
		\hline
		%% THIS IS OK FREE AND CONFINED
		$5_2 \rightarrow K$ & $0$ & $0.1$ & $0.11$ & $0.49$ & $0.26$ & $0.80$ \\ 
		%$5_2 \rightarrow 4_1$ & $0.02$ & $0$ & $0$ & $0$ & $0$ & $0$ \\ 
		$5_2 \rightarrow 3_1$ & $0.5$ & $0.25$ & $0.53$ & $0.2$ & $0.23$ & $0.13$ \\
		$5_2 \rightarrow 0_1$ & $0.5$ & $0.65$ & $0.36$ & $0.31$ & $0.51$ & $0.07$  \\
		\hline
		$5_1 \rightarrow K$ & $0$ & $0.06$ & $0.15$ & $0.69$ & -- & $0.8$ \\ 
		$5_1 \rightarrow 3_1$ & $1$ & $0.94$ & $0.85$ & $0.31$ & -- & $0.13$\\ 
		\hline
		%% NOW OK FREE AND CONFINED
		$4_1 \rightarrow K$ & $0$ & $0.04$ & $0.08$ & $0.16$ & -- & $0.84$ \\ 	 
		$4_1 \rightarrow 0_1$ & $1$ & $0.96$ & $0.92$ & $0.84$ & -- & $0.16$ \\ 	 
		\hline
		$3_1 \rightarrow K$ & $0$ & $0.15$ & $0.1$ & $0.22$ & $0.2$ & $0.87$\\ 	 
		$3_1 \rightarrow 0_1$ & $1$ & $0.85$ & $0.9$ & $0.78$ & $0.8$ & $0.13$\\ 	 
		\hline
	\end{tabular}
	\caption{Knot transition probabilities in different models. Topology simplification through the synergistic model proposed in this work is compared with RP (Ref.~\cite{Vazquez2007}) and HJ (Ref.~\cite{Burnier2007}) models. The confined case is compared with RP simulations performed in this work. $\mathcal{K}_1 \rightarrow K$ denotes transition to any knot $K$ with equal or larger minimal crossing number. (HJ, hooked juxtaposition; RP, Random Passage). }
	\label{table:tabSI}
\end{table*}

\section{Movies Captions}
Colour scheme: Dark-grey beads mark the polymer backbone. Light-grey beads the beads belonging to the extruded portion of the polymer. Blue beads the location of SMC heads. Green beads the location of TopoII.

\begin{enumerate}
\item Supplementary Movie M1: Localisation of a trefoil knot driven by a single processive SMC protein;

\item Supplementary Movie M2: Localisation of a Hopf link driven by a single processive SMC protein;

\item Supplementary Movie M3: Localisation and simplification of a trefoil knot via synergistic SMC-driven extrusion and Topo2-mediated strand crossing;

\item Supplementary Movie M4: Localisation of a trefoil knot under spherical confinement driven by a single processive SMC protein;

\item Supplementary Movie M5: Step-wise simplification of a $7_1$ knot under spherical confinement;

\item Supplementary Movie M6: Simultaneous extrusion of two parallel SMC proteins localise essential crossings but stabilise a delocalised knotted state (shortest knotted arc is shown in cyan); this state can be simplified by TopoII recruited by SMC.

\item Supplementary Movie M7: Unknotting of a trefoil by two diffusing SMC proteins (blue beads). At each time step one SMC is selected at random, and a TopoII (green beads) is located in front of its motion.

\end{enumerate}
\end{document}